\documentclass[longauth]{aa}  
\usepackage{graphicx}
\usepackage{txfonts}
\usepackage{natbib}
\usepackage{ulem}
                                
\usepackage[colorlinks=true,linkcolor=blue,filecolor=blue,urlcolor=blue,citecolor=blue]{hyperref}
\usepackage{amsmath} 
\usepackage{soul} 
\usepackage[switch]{linenoaa}
\begin{document} 

   \title{Very-high-energy observations of the Seyfert galaxy NGC~4151\\with MAGIC}

   \subtitle{Indication of another gamma-ray obscured candidate neutrino source}

%
\author{\tiny K.~Abe\inst{1} \and
S.~Abe\inst{2} \and
J.~Abhir\inst{3} \and
A.~Abhishek\inst{4} \and
V.~A.~Acciari\inst{5} \and
A.~Aguasca-Cabot\inst{6} \and
I.~Agudo\inst{7} \and
T.~Aniello\inst{8} \and
S.~Ansoldi\inst{9,43} \and
L.~A.~Antonelli\inst{8}
% ...
}
\institute { Japanese MAGIC Group: Department of Physics, Tokai University, Hiratsuka, 259-1292 Kanagawa, Japan
\and Japanese MAGIC Group: Institute for Cosmic Ray Research (ICRR), The University of Tokyo, Kashiwa, 277-8582 Chiba, Japan
\and ETH Z\"urich, CH-8093 Z\"urich, Switzerland
\and Universit\`a di Siena and INFN Pisa, I-53100 Siena, Italy
\and Institut de F\'isica d'Altes Energies (IFAE), The Barcelona Institute of Science and Technology (BIST), E-08193 Bellaterra (Barcelona), Spain
\and Universitat de Barcelona, ICCUB, IEEC-UB, E-08028 Barcelona, Spain
\and Instituto de Astrof\'isica de Andaluc\'ia-CSIC, Glorieta de la Astronom\'ia s/n, 18008, Granada, Spain
\and National Institute for Astrophysics (INAF), I-00136 Rome, Italy
\and Universit\`a di Udine and INFN Trieste, I-33100 Udine, Italy
\and Max-Planck-Institut f\"ur Physik, D-85748 Garching, Germany
\and Universit\`a di Padova and INFN, I-35131 Padova, Italy
\and Instituto de Astrof\'isica de Canarias and Dpto. de  Astrof\'isica, Universidad de La Laguna, E-38200, La Laguna, Tenerife, Spain
\and Croatian MAGIC Group: University of Zagreb, Faculty of Electrical Engineering and Computing (FER), 10000 Zagreb, Croatia
\and Saha Institute of Nuclear Physics, A CI of Homi Bhabha National Institute, Kolkata 700064, West Bengal, India
\and Centro Brasileiro de Pesquisas F\'isicas (CBPF), 22290-180 URCA, Rio de Janeiro (RJ), Brazil
\and IPARCOS Institute and EMFTEL Department, Universidad Complutense de Madrid, E-28040 Madrid, Spain
\and University of Lodz, Faculty of Physics and Applied Informatics, Department of Astrophysics, 90-236 Lodz, Poland
\and Centro de Investigaciones Energ\'eticas, Medioambientales y Tecnol\'ogicas, E-28040 Madrid, Spain
\and Departament de F\'isica, and CERES-IEEC, Universitat Aut\`onoma de Barcelona, E-08193 Bellaterra, Spain
\and Universit\`a di Pisa and INFN Pisa, I-56126 Pisa, Italy
\and INFN MAGIC Group: INFN Sezione di Bari and Dipartimento Interateneo di Fisica dell'Universit\`a e del Politecnico di Bari, I-70125 Bari, Italy
\and Department for Physics and Technology, University of Bergen, Norway
\and INFN MAGIC Group: INFN Sezione di Torino and Universit\`a degli Studi di Torino, I-10125 Torino, Italy
\and Croatian MAGIC Group: University of Rijeka, Faculty of Physics, 51000 Rijeka, Croatia
\and Universit\"at W\"urzburg, D-97074 W\"urzburg, Germany
\and Technische Universit\"at Dortmund, D-44221 Dortmund, Germany
\and Japanese MAGIC Group: Physics Program, Graduate School of Advanced Science and Engineering, Hiroshima University, 739-8526 Hiroshima, Japan
\and Deutsches Elektronen-Synchrotron (DESY), D-15738 Zeuthen, Germany
\and Armenian MAGIC Group: ICRANet-Armenia, 0019 Yerevan, Armenia
\and Croatian MAGIC Group: University of Split, Faculty of Electrical Engineering, Mechanical Engineering and Naval Architecture (FESB), 21000 Split, Croatia
\and Croatian MAGIC Group: Josip Juraj Strossmayer University of Osijek, Department of Physics, 31000 Osijek, Croatia
\and Finnish MAGIC Group: Finnish Centre for Astronomy with ESO, Department of Physics and Astronomy, University of Turku, FI-20014 Turku, Finland
\and University of Geneva, Chemin d'Ecogia 16, CH-1290 Versoix, Switzerland
\and Inst. for Nucl. Research and Nucl. Energy, Bulgarian Academy of Sciences, BG-1784 Sofia, Bulgaria
\and INFN MAGIC Group: INFN Sezione di Catania and Dipartimento di Fisica e Astronomia, University of Catania, I-95123 Catania, Italy
\and Japanese MAGIC Group: Department of Physics, Yamagata University, Yamagata 990-8560, Japan
\and Finnish MAGIC Group: Space Physics and Astronomy Research Unit, University of Oulu, FI-90014 Oulu, Finland
\and Japanese MAGIC Group: Chiba University, ICEHAP, 263-8522 Chiba, Japan
\and Japanese MAGIC Group: Institute for Space-Earth Environmental Research and Kobayashi-Maskawa Institute for the Origin of Particles and the Universe, Nagoya University, 464-6801 Nagoya, Japan
\and Japanese MAGIC Group: Department of Physics, Kyoto University, 606-8502 Kyoto, Japan
\and INFN MAGIC Group: INFN Roma Tor Vergata, I-00133 Roma, Italy
\and Japanese MAGIC Group: Department of Physics, Konan University, Kobe, Hyogo 658-8501, Japan
\and also at International Center for Relativistic Astrophysics (ICRA), Rome, Italy
% ...
}

   \date{Received XXX; accepted XXX}

  \abstract
{Seyfert galaxies are emerging as a promising source class of high-energy neutrinos. 
The Seyfert galaxies NGC~4151 and NGC~1068 have come up respectively
as the most promising counterparts of a 3$\sigma$ and of a 4.2$\sigma$ neutrino excesses detected by IceCube in the TeV energy range. 
Constraining the very-high-energy (VHE) emission associated with the neutrino signal is crucial to unveil the mechanism and site of neutrino production. 
In this work, we present the first results of the VHE observations ($\sim$29 hours) of NGC~4151 with the MAGIC telescopes. 
We detect no gamma-ray excess in the direction of NGC~4151, and we derive constraining upper limits on the VHE gamma-ray flux. The integral flux upper limit (at the 95$\%$ confidence level) above 200 GeV is $f = 2.3 \times 10^{-12}$ cm$^{-2}$ s$^{-1}$. 
The comparison of the MAGIC and IceCube measurements suggests the presence of a gamma-ray obscured accelerator, and it allows us to constrain the gamma-ray optical depth and the size of the neutrino production site.}


   \maketitle

\nolinenumbers
\section{Introduction} \label{intro}

After more than a decade from the IceCube discovery of an extra-terrestrial flux of high-energy neutrinos \citep{IceCube2013_Science}, the nature of the astrophysical sources responsible for the production of these cosmic messengers is still unknown. The spectral properties of the diffuse neutrino flux have been characterized with increasing accuracy in the energy range from a few TeV up to a few PeV \citep{IceCube-Diffuse}. 
The comparison of this flux with the isotropic gamma-ray background observed by Fermi-LAT \citep{Ackermann15} disfavors optically-thin gamma-ray emitters with soft spectra (spectral index $\gtrsim 2.2$) as the dominant neutrino source class, at least below 100 TeV \citep{Murase2013,Aartsen15,Bechtol17,Peretti20,Fang_2022}.
These results point towards the possible existence of gamma-ray obscured cosmic-ray accelerators \citep{Murase2016}.
The existence of such gamma-ray obscured powerful neutrino sources was already hypothesized by pioneering works from the late seventies \citep{Berezinsky_1977,Silberberg_1979,Eichler_1979,Berezinsky_1981,Berezinsky:1990qxi}. 
The nearest neighborhoods of active galactic nuclei (AGN) were one of the most promising environments for these conditions to be met ~\citep{Stecker1991,Stecker91-2}.
In the vicinity of the supermassive black hole \citep[SMBH, see e.g.][]{Padovani24} the high density of gas and radiation provides the right target for protons to sustain the production of TeV neutrinos, while the AGN photon field is also strong enough to efficiently absorb gamma rays through pair-production.

To date, indeed, the Seyfert 2 galaxy NGC~1068 has been found as the most significant hotspot in the Northern Sky in a recent IceCube analysis~\citep{Abbasi22}, and observations in the gamma-ray band indicate that the environment where the neutrinos are produced must be opaque to GeV-to-TeV gamma rays \citep{Aartsen20,Acciari19,Ajello23,Abbasi24}.
This was possible also thanks to the modest distance of the source, 10.1 Mpc \citep{Tully08,Padovani24}, for which the absorption of the extragalactic background light (EBL) is not a limitation \citep{Franceschini_EBL}.
The upper limits (ULs) on the very-high-energy (VHE) gamma-ray flux set by the Major Atmospheric Gamma Imaging Cherenkov (MAGIC) telescopes \citep{Acciari19} are more than one order of magnitude below the inferred neutrino flux. This suggests that the source is not optically thin to gamma rays. It is possible to reach this conclusion due to our understanding of the typical hadronic interaction channels, 
since hadronuclear ($pp$) and photo-hadronic ($p\gamma$) interactions produce comparable fluxes of gamma rays  and neutrinos as a result of similar production rates of neutral and charged pions.

How particles can be efficiently accelerated up to hundreds of TeV in the AGN vicinity  is not yet understood. Several mechanisms such as diffusive shock acceleration \citep{Inoue19,Inoue20,Peretti4151,Inoue2022}, stochastic acceleration \citep{Murase20,Eichmann20,Murase22,Murase24,Fiorillo24b,Lemoine:2024roa}, and magnetic reconnection \citep{Kheirandish21,Fiorillo24a,Mbarek24} have been proposed to explain the acceleration process. Regardless of the specific acceleration mechanism, it was shown that if energetic protons are injected into the AGN neighborhood, they are able to produce a neutrino emission at the level of flux observed by IceCube without exceeding the gamma-ray flux. Here gamma rays are efficiently absorbed and reprocessed via electromagnetic cascade reactions to the poorly explored MeV range \citep{Murase20}.

A key question we aim to address in this work is whether NGC~1068 is a peculiar source or representative of a broader population of gamma-ray obscured AGN neutrino sources.

Several acceleration models developed for the AGN nearest neighborhoods are based on equipartition arguments leading to a correlation between the neutrino luminosity and the X-ray luminosity of the AGN.
Consequently, several studies searching for neutrino signal in X-ray bright Seyfert galaxies have been carried out.
Interestingly, an independent analysis of the publicly available 10-year  IceCube dataset \citep{Neronov24}, along with IceCube analyses of extended datasets using different muon track event selections  \citep{Abbasi24,Abbasi24b}, 
have reported sizable neutrino excesses from some of the X-ray bright Seyfert galaxies in the  Swift-BAT catalog.
These findings reinforce the idea that Seyfert galaxies can be high-energy neutrino sources potentially dominating  the diffuse neutrino flux in the TeV range \citep{Padovani_24_Seyfert,Ambrosone24,Fiorillo25}.

Among them, the nearby Seyfert galaxy NGC~4151 \citep[$D = 15.8$ Mpc;][]{Yuan_distance} has been identified as a likely neutrino source. NGC~4151 is located at a distance of only 0.18$^{\circ}$ from the fourth most significant hotspot in the IceCube search for northern neutrino point sources \citep[see supplement of][]{Abbasi22}. \cite{Neronov24} reported a $\simeq$ 3$\sigma$ neutrino excess at the position of NGC~4151. \cite{Abbasi24} also identified NGC~4151 as one of the most interesting  candidates in their search for sources individually detectable by IceCube, 
reporting a post-trial significance of $\simeq$ 2.9 $\sigma$.

Although the neutrino excess from NGC~4151 is not yet statistically significant to claim a detection, it is interesting to consider the possibility that it corresponds to a real signal, and it may therefore represent a further indication of particle acceleration and hadronic interactions at work in a Seyfert galaxy. This motivates our dedicated search for VHE gamma-ray emission from this source.

NGC~4151 is classified as a Seyfert 1/1.5 \citep{Osterbrock76}.
The inner regions of Seyfert 1 are less obscured than Seyfert 2 providing a direct knowledge of the intrinsic AGN photon fields.
This allows for better constraints on neutrino emission and gamma-ray absorption compared to Compton-thick Seyfert 2 galaxies like NGC~1068. 
Investigating the gamma-ray/neutrino connection in NGC~4151 is also fundamental to assess whether the neutrino production region exhibits similar  gamma-ray opacity among the different Seyfert classes.

In this work, we present the first VHE observations of NGC~4151 with the MAGIC telescopes, we discuss the implications of our observations, and we derive constraints on the gamma-ray optical depth and size of the neutrino production site.


\section{Observations and data analysis} \label{Observations}

The VHE observations were performed by the  MAGIC telescopes, which consist of two 17-m diameter Imaging Atmospheric Cherenkov Telescopes (IACTs) located at an altitude of 2231 m above sea level, on the Canary Island of La Palma, Spain at the Roque de los Muchachos Observatory \citep{Aleksic16A}.
MAGIC has a typical energy resolution of about 16\%, and has an angular resolution at energies of a few hundred GeV of less than 0.07 degrees \citep{Aleksic16}. 
Under dark night conditions, and for zenith angles $<$30$^{\circ}$, MAGIC reaches a trigger energy threshold of $\sim$50 GeV, and a sensitivity above 220 GeV of 0.67$\% \pm$ 0.04$\%$ of the Crab Nebula flux in 50 hr of observations \citep{Aleksic16}. 

The telescopes with a field of view diameter of 3.5 degrees observed NGC~4151 under mostly dark observational conditions and good weather from 31st of March 2024 until 25th of June 2024. The zenith angle ($Z_d$) distribution of the data ranges from 10$^{\circ}$ and 50$^{\circ}$ with the majority of the data at $Z_d \lesssim$ 35$^{\circ}$. The analysis energy threshold determined by accounting for the zenith angle distribution  is $\sim$100 GeV.
In order to minimize systematics, the observations were performed in the wobble mode strategy \citep{Fomin94}, where the camera center is pointing at 0.4 degrees from the NGC~4151 position, so that the background and any possible gamma-ray signal can be estimated simultaneously.

After removing data affected by abnormally low background rates, and data with a median aerosol transmission at 9 km measured to be below 70$\%$ that of a clear night \citep{LIDAR_1, LIDAR_2}, the final sample consists of about 29 hours of effective observation time of good-quality data\footnote{93\% of the data have an aerosol transmission greater than 85\% of that of a clear night.} collected in 20 nights.
A small fraction of the observations (3.4 hours) are affected by the presence of the moon which increases the night sky background (NSB) fluctuations. However, the NSB level of the moon dataset is less than 2 times the one under dark conditions, so the moon and dark data samples were analyzed together applying the same standard image cleaning levels and Monte Carlo files. 

Data were analyzed using the standard MAGIC Analysis and Reconstruction Software (MARS) package \citep{Zanin13,Aleksic16}.
The data processing pipeline consists of calibration, image cleaning, and calculation of the second moment of the elliptical-shaped camera images of the telescopes,
the so-called Hillas parameterization. The random forest algorithm based on the reconstruction parameters is used to separate hadronic background from gamma-ray signal as well as to estimate the primary gamma-ray energy. To verify the validity of the reconstruction performance for NGC~4151, the same analysis pipeline has been applied to Crab Nebula data for similar periods and zenith angle distributions.

\section{Results} \label{Results}

\begin{figure}
    \centering
    \includegraphics[width=0.4\textwidth]{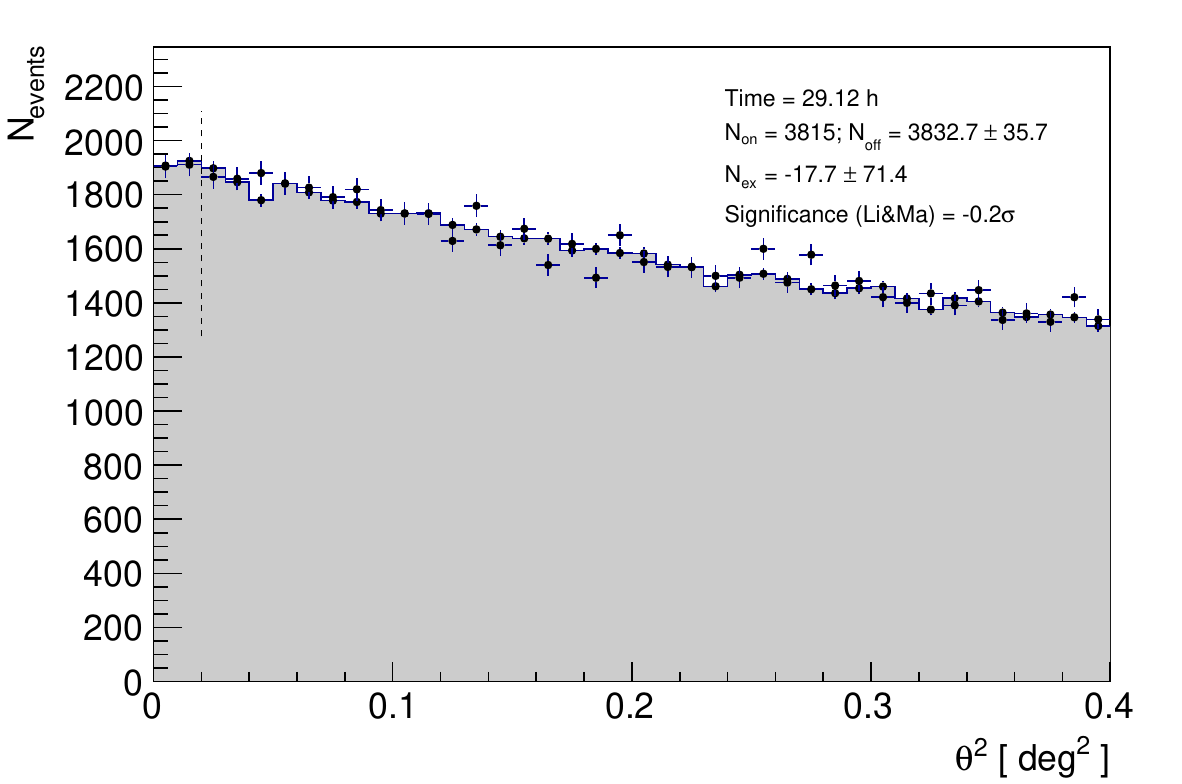}
    \caption{The squared angular distance ($\theta^{2}$) to the NGC~4151 position  distribution shows the number of gamma-ray candidate events as a function of the squared angular separation between the reconstructed gamma-ray direction and the NGC~4151 position. 
   The distribution of $\theta^{2}$, defined relative to the position of NGC~4151, 
 is indicated with the black points, while the gray histogram  corresponds to  the $\theta^{2}$  distribution defined relative to one of the background estimation points \citep{Fomin94}.The vertical dashed line defines the signal region within which the detection significance is computed.} 
    \label{odie_LE}
\end{figure}

\begin{figure}
    \centering
    \includegraphics[width=0.4\textwidth]{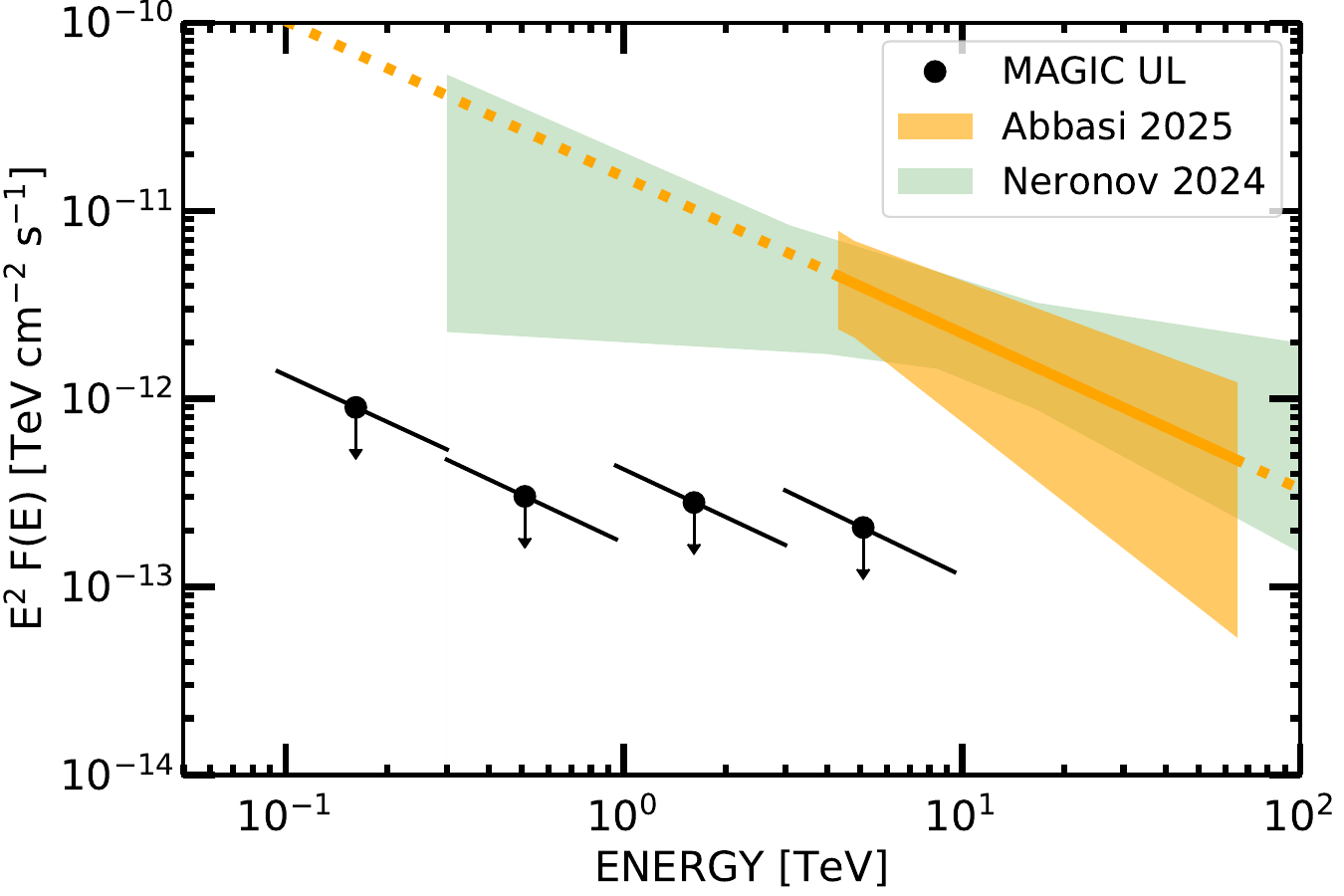}
    \caption{Multi-messenger SED of NGC 4151 in the VHE band. Black points indicate the MAGIC ULs derived in this work assuming a power-law spectral index $\Gamma$ = 2.83. The orange solid line and band refer to the best-fit and 1-$\sigma$ contour of the  single-flavor neutrino spectrum derived from IceCube data collected from muon tracks over 12 years \citep{Abbasi24}, while the dotted
orange corresponds to its extrapolation in energy. The green band refers to the single-flavor neutrino spectrum derived from the publicly available 10-year IceCube dataset \cite[adapted from][]{Neronov24}.}
    \label{NGC4151_MM_SED}
\end{figure}

Within the 29 hours of MAGIC observation, no significant gamma-ray emission was detected. 
Figure \ref{odie_LE} shows the number of detected gamma-ray candidate events as a function
of the squared angular distance to the NGC~4151 position (points) for energies greater than about 100 GeV, while the shaded gray region shows the estimated background.  
The vertical dashed line defines the boundary of the signal region within which the detection significance is computed.
We find 3815 gamma-like events for 3832.7$\pm$35.7 expected background events, which yields a significance of -0.2 $\sigma$ \citep{LiMa83}.
Given the lack of a significant signal from NGC~4151, ULs on the photon flux emitted in the VHE band were derived. Flux ULs are calculated following \cite{Rolke05}, with a confidence level of 95$\%$, and considering a systematic uncertainty on the gamma-ray detection efficiency of 30$\%$ \citep{Aleksic16}.\\
The integral flux UL (at the 95$\%$ confidence level) above 200 GeV is $f =$ 2.3$\times$10$^{-12}$ cm$^{-2}$ s$^{-1}$ assuming a power-law gamma-ray spectrum $\phi=\phi_0(E/E_0)^{-\Gamma}$, with spectral index $\Gamma$ = 2.83, while differential ULs up to $\sim$10 TeV are shown in Fig. \ref{NGC4151_MM_SED} and listed in Table \ref{tab1}.
Since our goal is to constrain the gamma-ray emission associated with the neutrino emission, the spectral index is assumed to be equal to the spectral index of the best-fit neutrino spectrum obtained from 12 years of IceCube (muon track) data \citep{Abbasi24}, which is reported in the same Figure (orange line) together with the 68\% statistical uncertainty.

In Figure \ref{NGC4151_MM_SED} we also show the neutrino spectrum derived by \cite{Neronov24} using the  publicly available 10 years of IceCube data (shaded green region). 
Although \cite{Abbasi24} determines that the central 68\% of the contribution to the test statistic for NGC 4151 comes from neutrinos with energies between 4.3 TeV and 65.2 TeV, we use the neutrino spectrum derived by \cite{Neronov24} to assess the impact of the energy range and flux normalization on our results.\\

\begin{table}
\caption{\label{tab1}Spectral Energy Distribution in the VHE band}
    \centering
    \resizebox{\columnwidth}{!}{%
    \begin{tabular}{ccccc}
    \hline
    \hline
      Energy   & Energy Low & Energy High   & 95$\%$ Flux UL \\
        (TeV)  & (TeV) & (TeV)  & (10$^{-13}$TeV cm$^{-2}$ s$^{-1}$) \\
       \hline
        0.162 (0.169) &  0.095 &  0.300 & 9.0 (8.9)\\
        0.510 (0.533) & 0.300 & 0.949 & 3.0 (3.2)\\
        1.613 (1.685) & 0.949 & 3.000 & 2.8 (3.1)\\
        5.103 (5.332) & 3.000 & 9.487 & 2.1 (2.4)\\
     \hline    
    \end{tabular}
    }
    \tablefoot{\tiny Differential flux ULs with a spectral index of $\Gamma$ = 2.83 ($\Gamma$ = 2.0) for four logarithmic energy bins between 0.095 and 9.5 TeV.  {\it First column:} average energy of the bin, reweighed according to the assumed spectral slope and the detected number of excesses in that bin \citep{Lafferty1995}. {\it Central columns:} lower and upper boundaries of the energy bins. {\it Last column:} flux ULs computed in each energy bin at a 95\% confidence level.}
\end{table}

\begin{figure}
    \centering
    \includegraphics[width=0.4\textwidth]{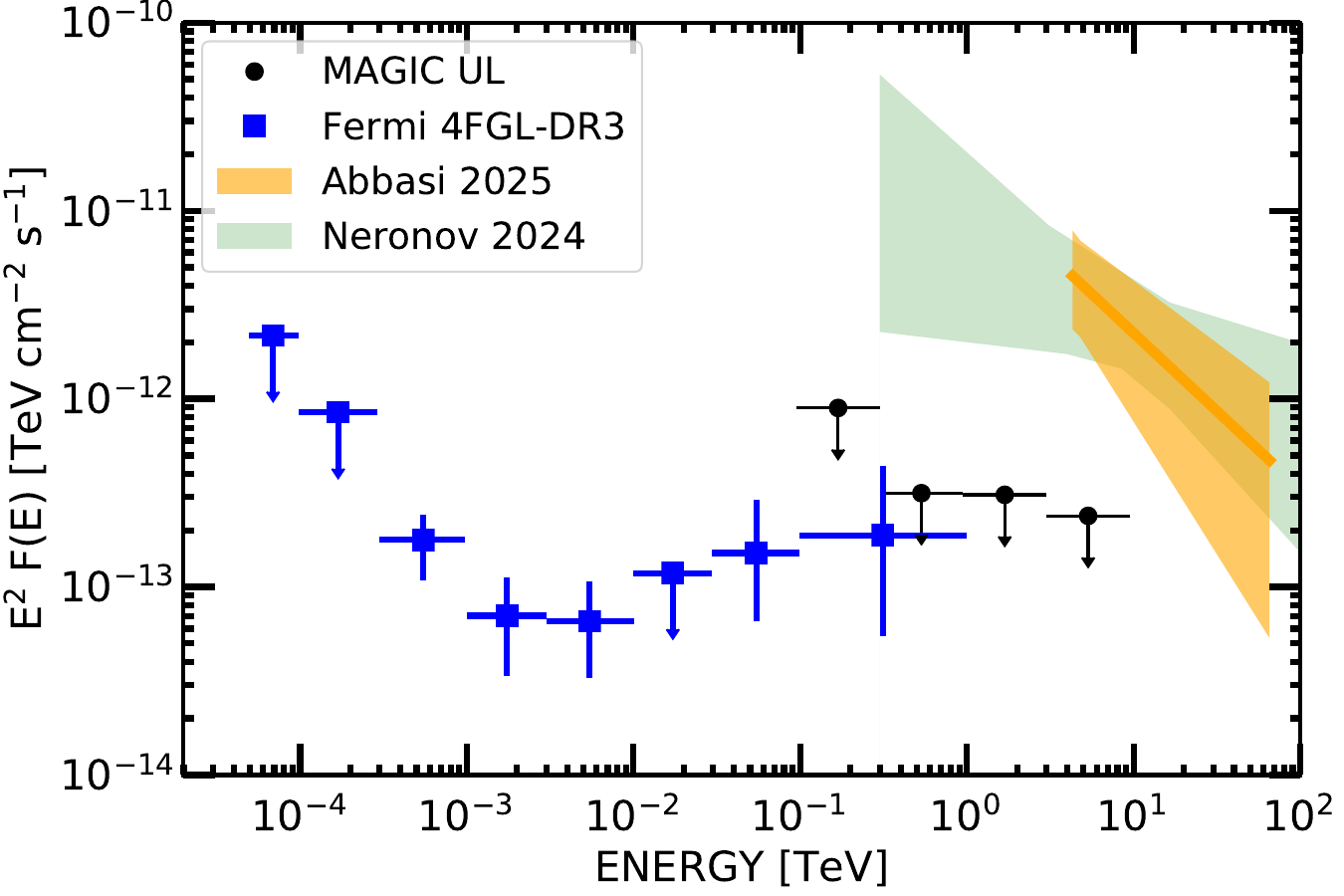}
    \caption{Multi-messenger SED of NGC 4151 in the HE and VHE bands. Black points indicate the MAGIC ULs obtained in this work assuming a power-law spectral index $\Gamma$ = 2.0. Blue data points indicate the Fermi-LAT spectrum of 4FGL J1210.3+3928 (spatially
coincident with NGC 4151)  from \cite{FermiDR3}. Neutrino spectra are the same as in Fig. \ref{NGC4151_MM_SED}.}
    \label{NGC4151_MM_SED_2}
\end{figure}

We also compute ULs assuming a standard power-law spectral index $\Gamma$ = 2.0, with the aim of allowing a direct comparison with similar sources already published in the literature \citep[e.g., NGC~1068;][]{Acciari19}.
These ULs, given  in Table \ref{tab1} and shown in Fig. \ref{NGC4151_MM_SED_2}, differ from those derived for $\Gamma = 2.83$ by at most  $\sim$10\%, indicating that they do not strongly depend on the assumed spectral index.
Figure \ref{NGC4151_MM_SED_2} extends the multi-messenger spectral energy distribution (SED) into the GeV band by including the spectral points detected by Fermi-LAT from the direction of NGC 4151.
In the Fermi-LAT band the region of NGC~4151 was first investigated in a study aimed at detecting the cumulative emission from AGN hosting ultra fast outflows \citep{Ajello_UFO}.  In that work, an individual significance of 4.2$\sigma$ was found from the position of NGC~4151.
More recent studies \citep{FermiDR3, fermilatdr4,Murase24,Peretti4151}  unveiled two point-like gamma-ray sources within 1 degree from the galaxy. One, 4FGL~J1211.6 + 3901 is associated to the $z\simeq0.6$ blazar FIRST J121134.2+390053, and is located at 0.4$^\circ$ from the center of NGC~4151. The other source, 4FGL~J1210.3+3928, is associated to the high synchrotron peak blazar 1E~1207.9+3945,  also located around $z =$ 0.61, and is found only 4.7 arcmin away from the center of NGC~4151. The presence of the latter background blazar makes the disentanglement among the potential emitters very challenging at the Fermi-LAT energies \citep[see e.g.][]{Murase24,Peretti4151}.
The blue data points in Figure \ref{NGC4151_MM_SED_2} indicate the gamma-ray spectrum of 4FGL~J1210.3+3928 from the 4FGL-DR3 source catalog \citep{FermiDR3}. 
We note that the ULs derived in the VHE band are in agreement with the Fermi-LAT data above $\sim$ 100 GeV, and that a possible contamination  of the gamma-ray emission from the background blazars could only affect the MAGIC observations at energies below $\sim$200 GeV, since at higher energies the EBL absorption coefficient is $\gtrsim$ 1 for a source emitting at a redshift of 0.6.
Since we do not detect the emission but only derive ULs on the gamma-ray flux, any contamination from the background blazar would make these limits even more conservative.

\section{Phenomenological constraints on the size of the neutrino source} \label{Discussion}

The non-detection of NGC 4151 in the VHE gamma-ray domain can aid in constraining some key properties of the neutrino source observed by IceCube in association with NGC 4151. 
For instance, the size of the emitting region can be probed through a model-independent study of its optical depth. 
Since NGC~4151, like NGC~1068, is expected to be a steady-state source of high-energy  neutrinos \citep{NGC1068_steady_1, NGC1068_steady_2}, we  work under the assumption of  steady production of neutrinos and gamma rays during the  observational time of IceCube and MAGIC. 
The spectra of neutrinos and gamma rays resulting from $pp$ interactions depend on the spectrum of primary protons, while in $p\gamma$ interactions, they are also determined by the spectrum of target photons. Neutrino production in a photon-rich environment also causes gamma-ray attenuation due to the onset of pair production in the ambient radiation field. The absorbed gamma-ray power is expected to be reprocessed at lower energies via the electromagnetic cascade. 
Thus, if the source is optically thick, the gamma-ray spectrum may undergo significant modifications before leaving the source, making the relation between observed gamma-ray and neutrino spectra more complex.
 However, the production spectra of gamma rays and neutrinos are expected to be similar as almost identical production rates of charged and neutral pions sharing the same energy are typical of hadronic and photo-hadronic interactions \citep{Kelner_pp,Kelner_pg,Roulet21,Condorelli25}.
Therefore, we assume that gamma rays are produced with the same power-law inferred for the neutrinos in the energy range observed by IceCube. 
Given $K$ as the average ratio of number of charged to neutral pions, the energy production rate of gamma rays and neutrinos produced in hadronic ($K =$ 2), or photo-hadronic ($K =$ 1) interactions, are related as follows~\citep{Murase2016}:
\begin{equation}
\label{Eq: neutrino-to-gamma}
    E^2_{\gamma} Q_{\gamma} (E_{\gamma}) \approx \frac{4}{3K} E^2_{\nu} Q_{\nu} (E_{\nu}) 
\end{equation}
where, $Q_{\gamma}$ ($Q_{\nu}$) is the production rate of gamma rays (all-flavor neutrinos) at energy $E_{\gamma}$ ($E_{\nu}\approx E_{\gamma}/2$).
This enables an estimation of the intrinsic gamma-ray emissivity given the neutrino one.
Equation \eqref{Eq: neutrino-to-gamma} depends on the interaction scenario and the slope of the particle distribution. To remain conservative, we adopt the most pessimistic assumption, which minimizes the gamma-to-neutrino production ratio, namely $K =$ 2 and a spectral slope $\Gamma = 2$. This choice ensures that our subsequent estimates are conservative and it corresponds to a gamma-ray flux approximately equal to twice the single-flavor neutrino flux at the same energy.

As the gamma-gamma absorption can play a dominant role in the AGN vicinity, it is fundamental to properly account for the AGN photon field.
To this end, we adopt the analytic prescriptions described in \cite{Marconi_2004} and \cite{Mullaney_2011} respectively for the optical to X-ray intrinsic AGN spectrum and the associated infrared component, and assume an X-ray luminosity of $L_X = 8 \times 10^{42} \, \rm erg \, s^{-1}$ \citep{Yang_2001,Gianolli_2023}.
The volume density of the higher energy background photons and gamma rays is set by assuming a specific value $R$ for the source radius. The low energy photon background and the associated optical depth are computed as reported in Appendix~\ref{App: Low energy}.

In our multi-messenger analysis we compute the observationally-motivated intrinsic gamma-ray luminosity as follows.
We assume as a reference the power-law neutrino spectrum provided by the IceCube collaboration in \cite{Abbasi24}, with best-fit normalization at 1 TeV $\phi_{\nu_{\mu}}^{1 \, \rm TeV}=$1.51$^{+0.99}_{-0.81}$ $\times$10$^{-11}$TeV $^{-1}$  cm$^{-2}$ s$^{-1}$,  and spectral slope $\Gamma$ = 2.83$^{+0.35}_{-0.28}$.
We assume the best fit value as a benchmark, and we also consider the whole $1\sigma$ uncertainty for both parameters $\phi_{\nu_{\mu}}^{\rm 1 TeV}$ and $\Gamma$. Since the low neutrino statistics prevent a precise determination of the IceCube spectrum, we extend the parameter exploration to include the case $\Gamma =$ 2, and examine how lower values of the spectral normalization affect our results.
We finally obtain the gamma-ray spectra following Eq.~\eqref{Eq: neutrino-to-gamma}.
The resulting gamma-ray luminosity (computed for the energy range $E \geq 300 \, \rm GeV$), $L_{\gamma}$, ranges from $10^{42}$ to $10^{43} \, \rm erg \, s^{-1}$, which is $\lesssim 5\%$ of the bolometric luminosity ($L_{\rm bol} \approx 1.9 \times 10^{44} \, \rm  erg \, s^{-1}$).

\begin{figure}
\centering
    \includegraphics[width=0.4\textwidth]{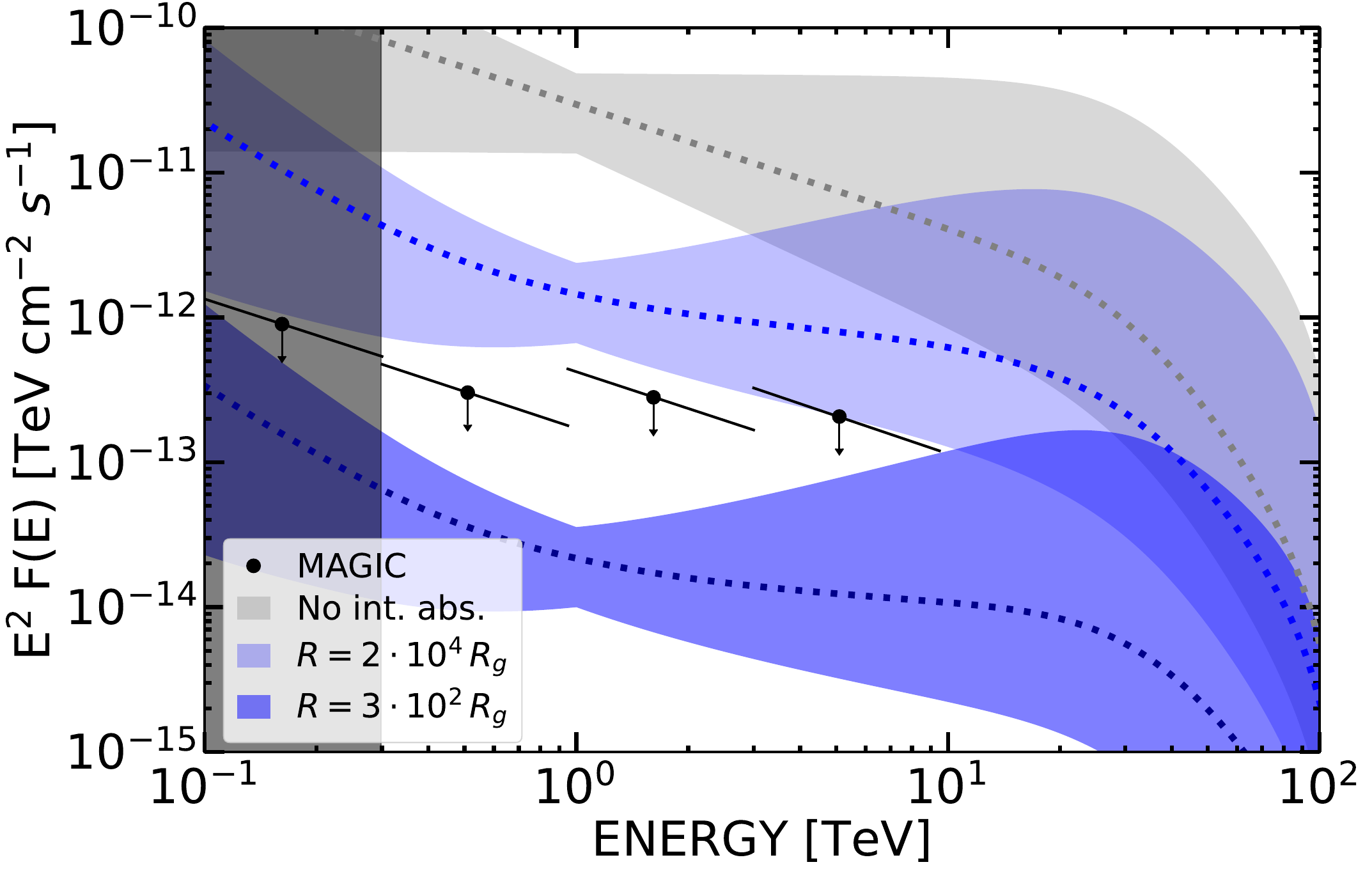}
    \caption{Phenomenological constraints on the size of the neutrino source obtained comparing the different predictions with MAGIC ULs (black points). 
    The gray dotted curve represents the unabsorbed gamma-ray flux corresponding to the best-fit neutrino spectrum, considering only the EBL absorption, while the gray band accounts for the  
    uncertainty in both $\phi_{\nu_{\mu}}^{\rm 1 TeV}$ and $\Gamma$.
    Similarly, the dotted light blue and blue curves, along with their respective shaded bands,
    represent the gamma-ray emission with different level of internal gamma-ray absorption, depending on the  assumed source radius: $2 \times 10^4$ (light blue) and $3 \times10^2$ (blue) gravitational radii. The dark gray area below 300 GeV is not taken into account in this analysis.}
    \label{Fig: phenomenology}
\end{figure}

Figure~\ref{Fig: phenomenology} illustrates the dependence of the gamma-ray spectrum on the assumed source size $R$ keeping constant only the impact of gamma-gamma absorption on the EBL. As discussed in Appendix \ref{App: Low energy}, the gamma-ray optical depth scales as $\tau_{\gamma \gamma} \sim R^{-1}$ thus the more compact the source the greater the gamma-ray optical depth.
First, we limit our analysis to the energy range probed by \cite{Abbasi24}, and then expand it to include the wider energy range considered in \cite{Neronov24}.
The comparison of the gamma-ray emission computed without internal absorption (gray line and band)
with  the MAGIC UL at 5 TeV clearly implies a substantial level of absorption inside the source.
Figure~\ref{Fig: phenomenology} also reveals that the gamma-ray flux is not compatible with the MAGIC UL for an emission size $R\gtrsim 2 \times 10^4 R_{g}$ (light blue band), where $R_{g}=GM_{\rm BH}/c^2$ is the gravitational radius. For a black hole mass of $M_{\rm BH}= 4.57 \times 10^7 \, M_{\odot}$ \citep{Bentz_2006,Gianolli_2023} this emission size corresponds to approximately  $ 4 \times 10^{-2} \, \rm pc$.
The largest radius allowing the gamma-ray spectral band to be completely localized  below the MAGIC UL is found at $R\simeq 3 \times 10^2 R_{g}$, which corresponds to $ 6 \times 10^{-4} \, \rm pc$.

Since we are also considering the analysis performed in \cite{Neronov24}, 
we extrapolate the reference neutrino spectrum of \cite{Abbasi24} and its 1$\sigma$ uncertainty to 300 GeV.
This allows for a multi-messenger analysis over an extended energy range probed by MAGIC observations.
As shown in Figure~\ref{Fig: phenomenology}, the derived constraints on the size of the neutrino production region remain valid. 
We note, however, that while the $1\sigma$ uncertainty band of \cite{Abbasi24} is in agreement with the flux uncertainty of \cite{Neronov24} for energies between 4 and 10 TeV, its extrapolation to lower energies does not match the lower bound of the neutrino flux found in their analysis (see Fig. \ref{NGC4151_MM_SED}). We verified that assuming this lower bound for the neutrino flux, our constraints on the neutrino production region are robust within a factor of $\sim$2.
We do not consider extrapolations of the gamma-ray spectra to energies lower than 300 GeV as they are not supported by the presence of neutrino data.
In addition, we notice that extrapolations to lower energies would also result in a gamma-ray luminosity comparable to or exceeding the AGN bolometric luminosity. 
Furthermore, such extrapolations would be better constrained by the emission of the Fermi-LAT source 4FGL~J1210.3+3928, which is spatially coincident with NGC~4151.

\section{Discussion} \label{MM_cons}

NGC~4151, similar to NGC~1068, appears as a bright TeV neutrino source candidate opaque in gamma rays  at the same energy. Therefore, the nearest neighborhood, $R\lesssim 10^{4} \, R_g$, of the accreting supermassive black hole \citep[often referred to as corona or cocoon, as discussed in pioneering works by][]{Berezinsky_1977,Berezinsky_1981,Berezinsky:1990qxi,Silberberg_1979,Eichler_1979,Stecker1991}, emerges as the most plausible candidate production site of such astronomical messengers \citep{Padovani24}.

The phenomenological test performed in the previous section is intended to illustrate the constraining power of MAGIC ULs to the size of the neutrino source. 
MAGIC observations are crucial for advancing our understanding of candidate neutrino sources such as Seyfert galaxies, since it operates in the VHE band, which partially overlaps with the operational energy range of the IceCube neutrino observatory. 
A model-independent constraint on the source’s opacity can only be obtained if a gamma-ray counterpart is not detected in the same energy range as that of the observed neutrinos.
The constraining power of other wavelengths remains model-dependent, as the production spectra of gamma rays and neutrinos can be very hard.

Therefore, focusing the analysis in the energy range above 4 TeV \citep[or 300 GeV when also][is considered]{Neronov24} is not a limitation as we do not intend to perform detailed physical modeling.
We only point out that a simple extrapolation of the gamma-ray flux down to the GeV range would lead to a total gamma-ray luminosity exceeding the bolometric one. 
However, we also observe that, a spectral break somewhere around TeV (or $10^2$ GeV) and a hard low-energy spectral slope ($\sim E^{-1}$) can be expected according to several physical models studying the nearest neighborhoods of AGN~\citep{Padovani24}.

In this context, the fiducial gamma-ray flux, if preceded by a hardening below  some energy break, might mimic the spectral behavior of models of cosmic-ray acceleration in the AGN corona through magnetic reconnection \citep{Kheirandish21,Fiorillo24a} or turbulence \citep{Murase20,Eichmann20,Murase22,Fiorillo24b,Mbarek24,Lemoine:2024roa}. 
Scenarios of diffusive shock acceleration at accretion shocks or at shocks developed in the early stages of a fast AGN-driven wind  \citep{Inoue19,Inoue20,Peretti4151,Inoue2022} are unlikely to produce spectra harder than $E^{-2}$, consequently their activity can be much better constrained by the Fermi-LAT observation in the GeV domain.\\

NGC~4151 is one of the brightest X-ray sources in the local Universe, and its possible association with neutrino emission aligns with predictions from acceleration models developed for the AGN nearest neighborhoods, which foresee a correlation between the X-ray and neutrino emission. However, this expectation is challenged by the fact that \cite{Abbasi24} reports no significant neutrino signal in the IceCube stacked analysis of the X-ray bright Swift-BAT sources. Nevertheless, uncertainties in intrinsic X-ray luminosity estimates, particularly for Compton-thick AGN, may introduce biases in the analysis, as strong absorption of the primary AGN emission and Compton reflection can significantly affect the determination of the X-ray luminosity.
On the other hand, gamma rays associated with neutrino emission, as outlined in the scenario presented here, are expected to be largely absorbed at energies above a few GeV (see Fig. \ref{Fig: Tau_gg}).
Future VHE observations, e.g., with the Cherenkov Telescope Array Observatory, will extend the spectral coverage and improve sensitivity, thereby providing
valuable constraints on the transport property of the emission region, its gamma-ray optical depth and the maximum energy energy of accelerated particles in these sources.


\section{Conclusions} \label{Conclusions}

NGC~4151 is recently in the spotlight as one of the brightest hotspots of the IceCube neutrino sky. Analyses using IceCube data identified NGC~4151 as a potential counterpart to a $\sim$3$\sigma$ neutrino excess in the TeV energy range.

The observation of the Seyfert galaxy NGC~4151  with the MAGIC telescopes resulted in a non-detection.  Data were accumulated from March to June 2024  for about 29 hours in good observational conditions. This allowed to set constraining ULs on the gamma-ray flux from this object in the energy range between 100 GeV and 10 TeV.

Gamma-ray and neutrino detectors operating in the same energy band are instrumental in obtaining model-independent constraints on the opacity of the neutrino production site. Since the neutrino spectrum in NGC~4151 spans the energy range $\sim$4-60 TeV \citep{Abbasi24}, the very-high-energy band is the optimal observational window to study absorption effects in this source.
The neutrino flux is observed at the level of $10^{-12}-10^{-11} \, \rm TeV \, cm^{-2} \, s^{-1}$ in the energy range  where MAGIC and IceCube overlap.
If this neutrino flux were confirmed, it would be incompatible with the ULs set by MAGIC in the context  of NGC~4151 as an optically thin gamma-ray source. 

The Seyfert 1/1.5 nature of NGC~4151 allows us to have a direct knowledge of the intrinsic X-ray luminosity of the AGN and consequently of the photon field of the AGN.
This allowed us to employ a phenomenological test and set an upper limit on the size of the neutrino production region below  $10^4$ gravitational radii.  
Such a result remains valid under different assumptions on the spectral shape of the injected particles.

This work showed that NGC~4151 is the second Seyfert galaxy, after NGC~1068, that might host a high-energy neutrino source optically thick to gamma rays. This suggests that, regardless of the Seyfert 1/2 classification, the neutrino production site in this class of sources has similar characteristics in terms of gamma-ray opacity.

To assess the validity of this interpretation, confirmation of the neutrino signal from this source and a more precise measurement of its spectrum are required. This will be achieved through the accumulation of additional data from current and next-generation neutrino detectors and more sensitive analyses with updated techniques.



\section*{Data availability}
The data underlying this article will be made available upon reasonable request to the corresponding authors.

\begin{acknowledgements}
A. Lamastra: project management, P.I. of MAGIC observations, MAGIC data analysis, theoretical interpretation, paper drafting; S. Mangano: MAGIC analysis cross-check, paper drafting; S. Menon:  MAGIC data analysis, paper drafting; E. Peretti: P.I. theory, theoretical modeling and interpretation, paper drafting; G. Peron: Fermi-LAT data interpretation and paper drafting; F. G. Saturni: theoretical interpretation and paper drafting.
The rest of the authors have contributed in one or several of the following ways: design, construction, maintenance and operation of the instrument(s) used to acquire the data; preparation and/or evaluation of the observation proposals; data acquisition, processing, calibration and/or reduction; production of analysis tools and/or related Monte Carlo simulations; overall discussions about the contents of the draft, as well as related refinements in the descriptions.

We would like to thank the Instituto de Astrof\'{\i}sica de Canarias for the excellent working conditions at the Observatorio del Roque de los Muchachos in La Palma. The financial support of the German BMBF, MPG and HGF; the Italian INFN and INAF; the Swiss National Fund SNF; the grants PID2019-104114RB-C31, PID2019-104114RB-C32, PID2019-104114RB-C33, PID2019-105510GB-C31, PID2019-107847RB-C41, PID2019-107847RB-C42, PID2019-107847RB-C44, PID2019-107988GB-C22, PID2022-136828NB-C41, PID2022-137810NB-C22, PID2022-138172NB-C41, PID2022-138172NB-C42, PID2022-138172NB-C43, PID2022-139117NB-C41, PID2022-139117NB-C42, PID2022-139117NB-C43, PID2022-139117NB-C44 funded by the Spanish MCIN/AEI/ 10.13039/501100011033 and “ERDF A way of making Europe”; the Indian Department of Atomic Energy; the Japanese ICRR, the University of Tokyo, JSPS, and MEXT; the Bulgarian Ministry of Education and Science, National RI Roadmap Project DO1-400/18.12.2020 and the Academy of Finland grant nr. 320045 is gratefully acknowledged. This work was also been supported by Centros de Excelencia ``Severo Ochoa'' y Unidades ``Mar\'{\i}a de Maeztu'' program of the Spanish MCIN/AEI/ 10.13039/501100011033 (CEX2019-000920-S, CEX2019-000918-M, CEX2021-001131-S) and by the CERCA institution and grants 2021SGR00426 and 2021SGR00773 of the Generalitat de Catalunya; by the Croatian Science Foundation (HrZZ) Project IP-2022-10-4595 and the University of Rijeka Project uniri-prirod-18-48; by the Deutsche Forschungsgemeinschaft (SFB1491) and by the Lamarr-Institute for Machine Learning and Artificial Intelligence; by the Polish Ministry Of Education and Science grant No. 2021/WK/08; and by the Brazilian MCTIC, CNPq and FAPERJ.
E.P. was supported by Agence Nationale de la Recherche (grant ANR-21-CE31-0028) and by INAF through “Assegni di ricerca per progetti di ricerca relativi a CTA e precursori”.
Reproduced with permission from Astronomy \& Astrophysics, \textcopyright\ ESO.
\end{acknowledgements}

\bibliographystyle{aa}
\bibliography{biblio}{}

\begin{thebibliography}{71}
\expandafter\ifx\csname natexlab\endcsname\relax\def\natexlab#1{#1}\fi

\bibitem[{{Aartsen} {et~al.}(2015){Aartsen}, {Abraham}, {Ackermann}, {Adams},
  {Aguilar}, {Ahlers}, {Ahrens}, {Altmann}, {Anderson}, {Archinger},
  {Arguelles}, {Arlen}, {Auffenberg}, {Bai}, {Barwick}, {Baum}, {Bay},
  {Beatty}, {Becker Tjus}, {Becker}, {Beiser}, {BenZvi}, {Berghaus}, {Berley},
  {Bernardini}, {Bernhard}, {Besson}, {Binder}, {Bindig}, {Bissok}, {Blaufuss},
  {Blumenthal}, {Boersma}, {Bohm}, {B{\"o}rner}, {Bos}, {Bose}, {B{\"o}ser},
  {Botner}, {Braun}, {Brayeur}, {Bretz}, {Brown}, {Buzinsky}, {Casey},
  {Casier}, {Cheung}, {Chirkin}, {Christov}, {Christy}, {Clark}, {Classen},
  {Coenders}, {Cowen}, {Cruz Silva}, {Daughhetee}, {Davis}, {Day}, {de
  Andr{\'e}}, {De Clercq}, {Dembinski}, {De Ridder}, {Desiati}, {de Vries}, {de
  Wasseige}, {de With}, {DeYoung}, {D{\'\i}az-V{\'e}lez}, {Dumm}, {Dunkman},
  {Eagan}, {Eberhardt}, {Ehrhardt}, {Eichmann}, {Euler}, {Evenson}, {Fadiran},
  {Fahey}, {Fazely}, {Fedynitch}, {Feintzeig}, {Felde}, {Filimonov}, {Finley},
  {Fischer-Wasels}, {Flis}, {Fuchs}, {Gaisser}, {Gaior}, {Gallagher},
  {Gerhardt}, {Ghorbani}, {Gier}, {Gladstone}, {Glagla}, {Gl{\"u}senkamp},
  {Goldschmidt}, {Golup}, {Gonzalez}, {Goodman}, {G{\'o}ra}, {Grant},
  {Gretskov}, {Groh}, {Gross}, {Ha}, {Haack}, {Haj Ismail}, {Hallgren},
  {Halzen}, {Hansmann}, {Hanson}, {Hebecker}, {Heereman}, {Helbing},
  {Hellauer}, {Hellwig}, {Hickford}, {Hignight}, {Hill}, {Hoffman}, {Hoffmann},
  {Holzapfel}, {Homeier}, {Hoshina}, {Huang}, {Huber}, {Huelsnitz}, {Hulth},
  {Hultqvist}, {In}, {Ishihara}, {Jacobi}, {Japaridze}, {Jero}, {Jurkovic},
  {Kaminsky}, {Kappes}, {Karg}, {Karle}, {Kauer}, {Keivani}, {Kelley}, {Kemp},
  {Kheirandish}, {Kiryluk}, {Kl{\"a}s}, {Klein}, {Kohnen}, {Kolanoski},
  {Konietz}, {Koob}, {K{\"o}pke}, {Kopper}, {Kopper}, {Koskinen}, {Kowalski},
  {Krings}, {Kroll}, {Kroll}, {Kunnen}, {Kurahashi}, {Kuwabara}, {Labare},
  {Lanfranchi}, {Larson}, {Lesiak-Bzdak}, {Leuermann}, {Leuner},
  {L{\"u}nemann}, {Madsen}, {Maggi}, {Mahn}, {Maruyama}, {Mase}, {Matis},
  {Maunu}, {McNally}, {Meagher}, {Medici}, {Meli}, {Menne}, {Merino}, {Meures},
  {Miarecki}, {Middell}, {Middlemas}, {Miller}, {Mohrmann}, {Montaruli},
  {Morse}, {Nahnhauer}, {Naumann}, {Niederhausen}, {Nowicki}, {Nygren},
  {Obertacke}, {Olivas}, {Omairat}, {O'Murchadha}, {Palczewski}, {Paul},
  {Pepper}, {P{\'e}rez de los Heros}, {Pfendner}, {Pieloth}, {Pinat},
  {Posselt}, {Price}, {Przybylski}, {P{\"u}tz}, {Quinnan}, {R{\"a}del},
  {Rameez}, {Rawlins}, {Redl}, {Reimann}, {Relich}, {Resconi}, {Rhode},
  {Richman}, {Richter}, {Riedel}, {Robertson}, {Rongen}, {Rott}, {Ruhe},
  {Ruzybayev}, {Ryckbosch}, {Saba}, {Sabbatini}, {Sander}, {Sandrock},
  {Sandroos}, {Sarkar}, {Schatto}, {Scheriau}, {Schimp}, {Schmidt}, {Schmitz},
  {Schoenen}, {Sch{\"o}neberg}, {Sch{\"o}nwald}, {Schukraft}, {Schulte},
  {Seckel}, {Seunarine}, {Shanidze}, {Smith}, {Soldin}, {Spiczak}, {Spiering},
  {Stahlberg}, {Stamatikos}, {Stanev}, {Stanisha}, {Stasik}, {Stezelberger},
  {Stokstad}, {St{\"o}ssl}, {Strahler}, {Str{\"o}m}, {Strotjohann}, {Sullivan},
  {Sutherland}, {Taavola}, {Taboada}, {Ter-Antonyan}, {Terliuk},
  {Te{\v{s}}i{\'c}}, {Tilav}, {Toale}, {Tobin}, {Tosi}, {Tselengidou}, {Unger},
  {Usner}, {Vallecorsa}, {Vandenbroucke}, {van Eijndhoven}, {Vanheule}, {van
  Santen}, {Veenkamp}, {Vehring}, {Voge}, {Vraeghe}, {Walck}, {Wallace},
  {Wallraff}, {Wandkowsky}, {Weaver}, {Wendt}, {Westerhoff}, {Whelan},
  {Whitehorn}, {Wichary}, {Wiebe}, {Wiebusch}, {Wille}, {Williams}, {Wissing},
  {Wolf}, {Wood}, {Woschnagg}, {Xu}, {Xu}, {Xu}, {Yanez}, {Yodh}, {Yoshida},
  {Zarzhitsky}, {Zoll}, \& {IceCube Collaboration}}]{Aartsen15}
{Aartsen}, M.~G., {Abraham}, K., {Ackermann}, M., {et~al.} 2015, \apj, 809, 98

\bibitem[{{Aartsen} {et~al.}(2020){Aartsen}, {Ackermann}, {Adams}, {Aguilar},
  {Ahlers}, {Ahrens}, {Alispach}, {Andeen}, {Anderson}, {Ansseau}, {Anton},
  {Arg{\"u}elles}, {Auffenberg}, {Axani}, {Backes}, {Bagherpour}, {Bai},
  {Balagopal}, {Barbano}, {Barwick}, {Bastian}, {Baum}, {Baur}, {Bay},
  {Beatty}, {Becker}, {Becker Tjus}, {BenZvi}, {Berley}, {Bernardini},
  {Besson}, {Binder}, {Bindig}, {Blaufuss}, {Blot}, {Bohm}, {B{\"o}rner},
  {B{\"o}ser}, {Botner}, {B{\"o}ttcher}, {Bourbeau}, {Bourbeau}, {Bradascio},
  {Braun}, {Bron}, {Brostean-Kaiser}, {Burgman}, {Buscher}, {Busse}, {Carver},
  {Chen}, {Cheung}, {Chirkin}, {Choi}, {Clark}, {Classen}, {Coleman}, {Collin},
  {Conrad}, {Coppin}, {Correa}, {Cowen}, {Cross}, {Dave}, {De Clercq},
  {DeLaunay}, {Dembinski}, {Deoskar}, {De Ridder}, {Desiati}, {de Vries}, {de
  Wasseige}, {de With}, {DeYoung}, {Diaz}, {D{\'\i}az-V{\'e}lez}, {Dujmovic},
  {Dunkman}, {Dvorak}, {Eberhardt}, {Ehrhardt}, {Eller}, {Engel}, {Evenson},
  {Fahey}, {Fazely}, {Felde}, {Filimonov}, {Finley}, {Fox}, {Franckowiak},
  {Friedman}, {Fritz}, {Gaisser}, {Gallagher}, {Ganster}, {Garrappa},
  {Gerhardt}, {Ghorbani}, {Glauch}, {Gl{\"u}senkamp}, {Goldschmidt},
  {Gonzalez}, {Grant}, {Griffith}, {Griswold}, {G{\"u}nder}, {G{\"u}nd{\"u}z},
  {Haack}, {Hallgren}, {Halliday}, {Halve}, {Halzen}, {Hanson}, {Haungs},
  {Hebecker}, {Heereman}, {Heix}, {Helbing}, {Hellauer}, {Henningsen},
  {Hickford}, {Hignight}, {Hill}, {Hoffman}, {Hoffmann}, {Hoinka},
  {Hokanson-Fasig}, {Hoshina}, {Huang}, {Huber}, {Huber}, {Hultqvist},
  {H{\"u}nnefeld}, {Hussain}, {In}, {Iovine}, {Ishihara}, {Japaridze}, {Jeong},
  {Jero}, {Jones}, {Jonske}, {Joppe}, {Kang}, {Kang}, {Kappes}, {Kappesser},
  {Karg}, {Karl}, {Karle}, {Katz}, {Kauer}, {Kelley}, {Kheirandish}, {Kim},
  {Kintscher}, {Kiryluk}, {Kittler}, {Klein}, {Koirala}, {Kolanoski},
  {K{\"o}pke}, {Kopper}, {Kopper}, {Koskinen}, {Kowalski}, {Krings},
  {Kr{\"u}ckl}, {Kulacz}, {Kurahashi}, {Kyriacou}, {Labare}, {Lanfranchi},
  {Larson}, {Lauber}, {Lazar}, {Leonard}, {Leszczy{\'n}ska}, {Leuermann},
  {Liu}, {Lohfink}, {Lozano Mariscal}, {Lu}, {Lucarelli}, {L{\"u}nemann},
  {Luszczak}, {Lyu}, {Ma}, {Madsen}, {Maggi}, {Mahn}, {Makino}, {Mallik},
  {Mallot}, {Mancina}, {Mari{\c{s}}}, {Maruyama}, {Mase}, {Matis}, {Maunu},
  {McNally}, {Meagher}, {Medici}, {Medina}, {Meier}, {Meighen-Berger}, {Menne},
  {Merino}, {Meures}, {Micallef}, {Mockler}, {Moment{\'e}}, {Montaruli},
  {Moore}, {Morse}, {Moulai}, {Muth}, {Nagai}, {Naumann}, {Neer},
  {Niederhausen}, {Nisa}, {Nowicki}, {Nygren}, {Obertacke Pollmann}, {Oehler},
  {Olivas}, {O'Murchadha}, {O'Sullivan}, {Palczewski}, {Pandya}, {Pankova},
  {Park}, {Peiffer}, {P{\'e}rez de los Heros}, {Philippen}, {Pieloth}, {Pinat},
  {Pizzuto}, {Plum}, {Porcelli}, {Price}, {Przybylski}, {Raab}, {Raissi},
  {Rameez}, {Rauch}, {Rawlins}, {Rea}, {Reimann}, {Relethford}, {Renschler},
  {Renzi}, {Resconi}, {Rhode}, {Richman}, {Robertson}, {Rongen}, {Rott},
  {Ruhe}, {Ryckbosch}, {Rysewyk}, {Safa}, {Sanchez Herrera}, {Sandrock},
  {Sandroos}, {Santander}, {Sarkar}, {Sarkar}, {Satalecka}, {Schaufel},
  {Schieler}, {Schlunder}, {Schmidt}, {Schneider}, {Schneider}, {Schr{\"o}der},
  {Schumacher}, {Sclafani}, {Seckel}, {Seunarine}, {Shefali}, {Silva},
  {Snihur}, {Soedingrekso}, {Soldin}, {Song}, {Spiczak}, {Spiering},
  {Stachurska}, {Stamatikos}, {Stanev}, {Stein}, {Steinm{\"u}ller}, {Stettner},
  {Steuer}, {Stezelberger}, {Stokstad}, {St{\"o}{\ss}l}, {Strotjohann},
  {St{\"u}rwald}, {Stuttard}, {Sullivan}, {Taboada}, {Tenholt}, {Ter-Antonyan},
  {Terliuk}, {Tilav}, {Tollefson}, {Tomankova}, {T{\"o}nnis}, {Toscano},
  {Tosi}, {Trettin}, {Tselengidou}, {Tung}, {Turcati}, {Turcotte}, {Turley},
  {Ty}, {Unger}, {Unland Elorrieta}, {Usner}, {Vandenbroucke}, {Van Driessche},
  {van Eijk}, {van Eijndhoven}, {Vanheule}, {van Santen}, {Vraeghe}, {Walck},
  {Wallace}, {Wallraff}, {Wandkowsky}, {Watson}, {Weaver}, {Weindl}, {Weiss},
  {Weldert}, {Wendt}, {Werthebach}, {Whelan}, {Whitehorn}, {Wiebe}, {Wiebusch},
  {Wille}, {Williams}, {Wills}, {Wolf}, {Wood}, {Wood}, {Woschnagg}, {Wrede},
  {Xu}, {Xu}, {Xu}, {Yanez}, {Yodh}, {Yoshida}, {Yuan}, \&
  {Z{\"o}cklein}}]{Aartsen20}
{Aartsen}, M.~G., {Ackermann}, M., {Adams}, J., {et~al.} 2020, \prl, 124,
  051103

\bibitem[{{Abbasi} {et~al.}(2024{\natexlab{a}}){Abbasi}, {Ackermann}, {Adams},
  {Agarwalla}, {Aguilar}, {Ahlers}, {Alameddine}, {Amin}, {Andeen}, {Anton},
  {Arg{\"u}elles}, {Ashida}, {Athanasiadou}, {Ausborm}, {Axani}, {Bai},
  {Balagopal}, {Baricevic}, {Barwick}, {Bash}, {Basu}, {Bay}, {Beatty}, {Becker
  Tjus}, {Beise}, {Bellenghi}, {Benning}, {BenZvi}, {Berley}, {Bernardini},
  {Besson}, {Blaufuss}, {Blot}, {Bontempo}, {Book}, {Boscolo Meneguolo},
  {B{\"o}ser}, {Botner}, {B{\"o}ttcher}, {Braun}, {Brinson}, {Brostean-Kaiser},
  {Brusa}, {Burley}, {Busse}, {Butterfield}, {Campana}, {Caracas}, {Carloni},
  {Carpio}, {Chattopadhyay}, {Chau}, {Chen}, {Chirkin}, {Choi}, {Clark},
  {Coleman}, {Collin}, {Connolly}, {Conrad}, {Coppin}, {Corley}, {Correa},
  {Cowen}, {Dave}, {De Clercq}, {DeLaunay}, {Delgado}, {Deng}, {Deoskar},
  {Desai}, {Desiati}, {de Vries}, {de Wasseige}, {DeYoung}, {Diaz},
  {D{\'\i}az-V{\'e}lez}, {Dittmer}, {Domi}, {Draper}, {Dujmovic}, {DuVernois},
  {Ehrhardt}, {Eidenschink}, {Eimer}, {Eller}, {Ellinger}, {El Mentawi},
  {Els{\"a}sser}, {Engel}, {Erpenbeck}, {Evans}, {Evenson}, {Fan}, {Fang},
  {Farrag}, {Fazely}, {Fedynitch}, {Feigl}, {Fiedlschuster}, {Finley},
  {Fischer}, {Fox}, {Franckowiak}, {F{\"u}rst}, {Gallagher}, {Ganster},
  {Garcia}, {Genton}, {Gerhardt}, {Ghadimi}, {Girard-Carillo}, {Glaser},
  {Gl{\"u}senkamp}, {Gonzalez}, {Goswami}, {Granados}, {Grant}, {Gray},
  {Gries}, {Griffin}, {Griswold}, {Groth}, {G{\"u}nther}, {Gutjahr}, {Ha},
  {Haack}, {Hallgren}, {Halliday}, {Halve}, {Halzen}, {Hamdaoui}, {Ha Minh},
  {Handt}, {Hanson}, {Hardin}, {Harnisch}, {Hatch}, {Haungs},
  {H{\"a}u{\ss}ler}, {Helbing}, {Hellrung}, {Hermannsgabner}, {Heuermann},
  {Heyer}, {Hickford}, {Hidvegi}, {Hill}, {Hill}, {Hoffman}, {Hori}, {Hoshina},
  {Hostert}, {Hou}, {Huber}, {Hultqvist}, {H{\"u}nnefeld}, {Hussain}, {Hymon},
  {Ishihara}, {Iwakiri}, {Jacquart}, {Janik}, {Jansson}, {Japaridze}, {Jeong},
  {Jin}, {Jones}, {Kamp}, {Kang}, {Kang}, {Kang}, {Kappes}, {Kappesser},
  {Kardum}, {Karg}, {Karl}, {Karle}, {Katil}, {Katz}, {Kauer}, {Kelley},
  {Khanal}, {Khatee Zathul}, {Kheirandish}, {Kiryluk}, {Klein}, {Kochocki},
  {Koirala}, {Kolanoski}, {Kontrimas}, {K{\"o}pke}, {Kopper}, {Koskinen},
  {Koundal}, {Kovacevich}, {Kowalski}, {Kozynets}, {Krishnamoorthi},
  {Kruiswijk}, {Krupczak}, {Kumar}, {Kun}, {Kurahashi}, {Lad}, {Lagunas
  Gualda}, {Lamoureux}, {Larson}, {Latseva}, {Lauber}, {Lazar}, {Lee}, {Leonard
  DeHolton}, {Leszczy{\'n}ska}, {Liao}, {Lincetto}, {Liubarska}, {Lohfink},
  {Love}, {Lozano Mariscal}, {Lu}, {Lucarelli}, {Luszczak}, {Lyu}, {Madsen},
  {Magnus}, {Mahn}, {Makino}, {Manao}, {Mancina}, {Marie Sainte},
  {Mari{\c{s}}}, {Marka}, {Marka}, {Marsee}, {Martinez-Soler}, {Maruyama},
  {Mayhew}, {McElroy}, {McNally}, {Mead}, {Meagher}, {Mechbal}, {Medina},
  {Meier}, {Merckx}, {Merten}, {Micallef}, {Mitchell}, {Montaruli}, {Moore},
  {Morii}, {Morse}, {Moulai}, {Mukherjee}, {Naab}, {Nagai}, {Nakos}, {Naumann},
  {Necker}, {Negi}, {Neumann}, {Niederhausen}, {Nisa}, {Noell}, {Novikov},
  {Nowicki}, {Obertacke Pollmann}, {O'Dell}, {Oeyen}, {Olivas}, {Orsoe},
  {Osborn}, {O'Sullivan}, {Pandya}, {Park}, {Parker}, {Paudel}, {Paul},
  {P{\'e}rez de los Heros}, {Pernice}, {Peterson}, {Philippen}, {Pizzuto},
  {Plum}, {Pont{\'e}n}, {Popovych}, {Prado Rodriguez}, {Pries},
  {Procter-Murphy}, {Przybylski}, {Raab}, {Rack-Helleis}, {Rawlins}, {Rechav},
  {Rehman}, {Reichherzer}, {Resconi}, {Reusch}, {Rhode}, {Riedel}, {Rifaie},
  {Roberts}, {Robertson}, {Rodan}, {Roellinghoff}, {Rongen}, {Rosted}, {Rott},
  {Ruhe}, {Ruohan}, {Ryckbosch}, {Safa}, {Saffer}, {Salazar-Gallegos},
  {Sampathkumar}, {Sandrock}, {Santander}, {Sarkar}, {Sarkar}, {Savelberg},
  {Savina}, {Schaile}, {Schaufel}, {Schieler}, {Schindler}, {Schl{\"u}ter},
  {Schl{\"u}ter}, {Schmeisser}, {Schmidt}, {Schneider}, {Schr{\"o}der},
  {Schumacher}, {Sclafani}, {Seckel}, {Seikh}, {Seo}, {Seunarine}, {Sevle
  Myhr}, {Shah}, {Shefali}, {Shimizu}, {Silva}, {Skrzypek}, {Smithers},
  {Snihur}, {Soedingrekso}, {S{\o}gaard}, {Soldin}, {Soldin}, {Sommani},
  {Spannfellner}, {Spiczak}, {Spiering}, {Stamatikos}, {Stanev},
  {Stezelberger}, {St{\"u}rwald}, {Stuttard}, {Sullivan}, {Taboada},
  {Ter-Antonyan}, {Terliuk}, {Thiesmeyer}, {Thompson}, {Thwaites}, {Tilav},
  {Tollefson}, {T{\"o}nnis}, {Toscano}, {Tosi}, {Trettin}, {Turcotte},
  {Twagirayezu}, {Unland Elorrieta}, {Upadhyay}, {Upshaw}, {Vaidyanathan},
  {Valtonen-Mattila}, {Vandenbroucke}, {van Eijndhoven}, {Vannerom}, {van
  Santen}, {Vara}, {Veitch-Michaelis}, {Venugopal}, {Vereecken}, {Verpoest},
  {Veske}, {Vijai}, {Walck}, {Wang}, {Weaver}, {Weigel}, {Weindl}, {Weldert},
  {Wen}, {Wendt}, {Werthebach}, {Weyrauch}, {Whitehorn}, {Wiebusch},
  {Williams}, {Witthaus}, {Wolf}, {Wolf}, {Wrede}, {Xu}, {Yanez}, {Yildizci},
  {Yoshida}, {Young}, {Yu}, {Yuan}, {Zhang}, {Zhelnin}, {Zilberman},
  {Zimmerman}, \& {IceCube Collaboration}}]{IceCube-Diffuse}
{Abbasi}, R., {Ackermann}, M., {Adams}, J., {et~al.} 2024{\natexlab{a}}, \prd,
  110, 022001

\bibitem[{{Abbasi} {et~al.}(2025){Abbasi}, {Ackermann}, {Adams}, {Agarwalla},
  {Aguilar}, {Ahlers}, {Alameddine}, {Amin}, {Andeen}, {Arg{\"u}elles},
  {Ashida}, {Athanasiadou}, {Ausborm}, {Axani}, {Bai}, {Balagopal V.},
  {Baricevic}, {Barwick}, {Bash}, {Basu}, {Bay}, {Beatty}, {Becker Tjus},
  {Beise}, {Bellenghi}, {Benning}, {BenZvi}, {Berley}, {Bernardini}, {Besson},
  {Blaufuss}, {Bloom}, {Blot}, {Bontempo}, {Book Motzkin}, {Boscolo Meneguolo},
  {B{\"o}ser}, {Botner}, {B{\"o}ttcher}, {Braun}, {Brinson}, {Brostean-Kaiser},
  {Brusa}, {Burley}, {Butterfield}, {Campana}, {Caracas}, {Carloni}, {Carpio},
  {Chattopadhyay}, {Chau}, {Chen}, {Chirkin}, {Choi}, {Clark}, {Coleman},
  {Collin}, {Connolly}, {Conrad}, {Coppin}, {Corley}, {Correa}, {Cowen},
  {Dave}, {De Clercq}, {DeLaunay}, {Delgado}, {Deng}, {Desai}, {Desiati}, {de
  Vries}, {de Wasseige}, {DeYoung}, {Diaz}, {D{\'\i}az-V{\'e}lez}, {Dierichs},
  {Dittmer}, {Domi}, {Draper}, {Dujmovic}, {Dutta}, {DuVernois}, {Ehrhardt},
  {Eidenschink}, {Eimer}, {Eller}, {Ellinger}, {El Mentawi}, {Els{\"a}sser},
  {Engel}, {Erpenbeck}, {Evans}, {Evenson}, {Fan}, {Fang}, {Farrag}, {Fazely},
  {Fedynitch}, {Feigl}, {Fiedlschuster}, {Finley}, {Fischer}, {Fox},
  {Franckowiak}, {Fukami}, {F{\"u}rst}, {Gallagher}, {Ganster}, {Garcia},
  {Garcia}, {Garg}, {Genton}, {Gerhardt}, {Ghadimi}, {Girard-Carillo},
  {Glaser}, {Gl{\"u}senkamp}, {Gonzalez}, {Goswami}, {Granados}, {Grant},
  {Gray}, {Gries}, {Griffin}, {Griswold}, {Groth}, {G{\"u}nther}, {Gutjahr},
  {Ha}, {Haack}, {Hallgren}, {Halve}, {Halzen}, {Hamdaoui}, {Ha Minh}, {Handt},
  {Hanson}, {Hardin}, {Harnisch}, {Hatch}, {Haungs}, {H{\"a}ussler}, {Helbing},
  {Hellrung}, {Hermannsgabner}, {Heuermann}, {Heyer}, {Hickford}, {Hidvegi},
  {Hill}, {Hill}, {Hoffman}, {Hori}, {Hoshina}, {Hostert}, {Hou}, {Huber},
  {Hultqvist}, {H{\"u}nnefeld}, {Hussain}, {Hymon}, {Ishihara}, {Iwakiri},
  {Jacquart}, {Janik}, {Jansson}, {Japaridze}, {Jeong}, {Jin}, {Jones}, {Kamp},
  {Kang}, {Kang}, {Kang}, {Kappes}, {Kappesser}, {Kardum}, {Karg}, {Karl},
  {Karle}, {Katil}, {Katz}, {Kauer}, {Kelley}, {Khanal}, {Khatee Zathul},
  {Kheirandish}, {Kiryluk}, {Klein}, {Kochocki}, {Koirala}, {Kolanoski},
  {Kontrimas}, {K{\"o}pke}, {Kopper}, {Koskinen}, {Koundal}, {Kovacevich},
  {Kowalski}, \& {Kozynets}}]{Abbasi24}
{Abbasi}, R., {Ackermann}, M., {Adams}, J., {et~al.} 2025, \apj, 981, 131

\bibitem[{{Abbasi} {et~al.}(2024{\natexlab{b}}){Abbasi}, {Ackermann}, {Adams},
  {Agarwalla}, {Aguilar}, {Ahlers}, {Alameddine}, {Amin}, {Andeen},
  {Arg{\"u}elles}, {Ashida}, {Athanasiadou}, {Ausborm}, {Axani}, {Bai},
  {Balagopal V.}, {Baricevic}, {Barwick}, {Bash}, {Basu}, {Bay}, {Beatty},
  {Becker Tjus}, {Beise}, {Bellenghi}, {Benning}, {BenZvi}, {Berley},
  {Bernardini}, {Besson}, {Blaufuss}, {Bloom}, {Blot}, {Bontempo}, {Book
  Motzkin}, {Boscolo Meneguolo}, {B{\"o}ser}, {Botner}, {B{\"o}ttcher},
  {Braun}, {Brinson}, {Brostean-Kaiser}, {Brusa}, {Burley}, {Butterfield},
  {Campana}, {Caracas}, {Carloni}, {Carpio}, {Chattopadhyay}, {Chau}, {Chen},
  {Chirkin}, {Choi}, {Clark}, {Coleman}, {Collin}, {Connolly}, {Conrad},
  {Coppin}, {Corley}, {Correa}, {Cowen}, {Dave}, {De Clercq}, {DeLaunay},
  {Delgado}, {Deng}, {Desai}, {Desiati}, {de Vries}, {de Wasseige}, {DeYoung},
  {Diaz}, {D{\'\i}az-V{\'e}lez}, {Dierichs}, {Dittmer}, {Domi}, {Draper},
  {Dujmovic}, {Dutta}, {DuVernois}, {Ehrhardt}, {Eidenschink}, {Eimer},
  {Eller}, {Ellinger}, {El Mentawi}, {Els{\"a}sser}, {Engel}, {Erpenbeck},
  {Evans}, {Evenson}, {Fan}, {Fang}, {Farrag}, {Fazely}, {Fedynitch}, {Feigl},
  {Fiedlschuster}, {Finley}, {Fischer}, {Fox}, {Franckowiak}, {Fukami},
  {F{\"u}rst}, {Gallagher}, {Ganster}, {Garcia}, {Garcia}, {Garg}, {Genton},
  {Gerhardt}, {Ghadimi}, {Girard-Carillo}, {Glaser}, {Glauch},
  {Gl{\"u}senkamp}, {Gonzalez}, {Goswami}, {Granados}, {Grant}, {Gray},
  {Gries}, {Griffin}, {Griswold}, {Groth}, {G{\"u}nther}, {Gutjahr}, {Ha},
  {Haack}, {Hallgren}, {Halve}, {Halzen}, {Hamdaoui}, {Minh}, {Handt},
  {Hanson}, {Hardin}, {Harnisch}, {Hatch}, {Haungs}, {H{\"a}u{\ss}ler},
  {Helbing}, {Hellrung}, {Hermannsgabner}, {Heuermann}, {Heyer}, {Hickford},
  {Hidvegi}, {Hill}, {Hill}, {Hoffman}, {Hori}, {Hoshina}, {Hostert}, {Hou},
  {Huber}, {Hultqvist}, {H{\"u}nnefeld}, {Hussain}, {Hymon}, {Ishihara},
  {Iwakiri}, {Jacquart}, {Janik}, {Jansson}, {Japaridze}, {Jeong}, {Jin},
  {Jones}, {Kamp}, {Kang}, {Kang}, {Kang}, {Kappes}, {Kappesser}, {Kardum},
  {Karg}, {Karl}, {Karle}, {Katil}, {Katz}, {Kauer}, {Kelley}, {Khanal},
  {Khatee Zathul}, {Kheirandish}, {Kiryluk}, {Klein}, {Kochocki}, {Koirala},
  {Kolanoski}, {Kontrimas}, {K{\"o}pke}, {Kopper}, {Koskinen}, {Koundal},
  {Kovacevich}, {Kowalski}, {Kozynets}, {Krishnamoorthi}, {Kruiswijk},
  {Krupczak}, {Kumar}, {Kun}, {Kurahashi}, {Lad}, {Lagunas Gualda},
  {Lamoureux}, {Larson}, {Latseva}, {Lauber}, {Lazar}, {Lee}, {DeHolton},
  {Leszczy{\'n}ska}, {Liao}, {Lincetto}, {Liu}, {Liu}, {Liubarska}, {Lohfink},
  {Love}, {Lozano Mariscal}, {Lu}, {Lucarelli}, {Luszczak}, {Lyu}, {Madsen},
  {Magnus}, {Mahn}, {Makino}, {Manao}, {Mancina}, {Sainte}, {Mari{\c{s}}},
  {Marka}, {Marka}, {Marsee}, {Martinez-Soler}, {Maruyama}, {Mayhew},
  {McNally}, {Mead}, {Meagher}, {Mechbal}, {Medina}, {Meier}, {Merckx},
  {Merten}, {Micallef}, {Mitchell}, {Montaruli}, {Moore}, {Morii}, {Morse},
  {Moulai}, {Mukherjee}, {Naab}, {Nagai}, {Nakos}, {Naumann}, {Necker}, {Negi},
  {Neste}, {Neumann}, {Niederhausen}, {Nisa}, {Noda}, {Noell}, {Novikov},
  {Obertacke Pollmann}, {O'Dell}, {Oeyen}, {Olivas}, {Orsoe}, {Osborn},
  {O'Sullivan}, {Pandya}, {Park}, {Parker}, {Paudel}, {Paul}, {P{\'e}rez de los
  Heros}, {Pernice}, {Peterson}, {Philippen}, {Pizzuto}, {Plum}, {Pont{\'e}n},
  {Popovych}, {Prado Rodriguez}, {Pries}, {Procter-Murphy}, {Przybylski},
  {Raab}, {Rack-Helleis}, {Ravn}, {Rawlins}, {Rechav}, {Rehman}, {Reichherzer},
  {Resconi}, {Reusch}, {Rhode}, {Riedel}, {Rifaie}, {Roberts}, {Robertson},
  {Rodan}, {Roellinghoff}, {Rongen}, {Rosted}, {Rott}, {Ruhe}, {Ruohan},
  {Ryckbosch}, {Safa}, {Saffer}, {Salazar-Gallegos}, {Sampathkumar},
  {Sandrock}, {Santander}, {Sarkar}, {Sarkar}, {Savelberg}, {Savina},
  {Schaile}, {Schaufel}, {Schieler}, {Schindler}, {Schl{\"u}ter},
  {Schl{\"u}ter}, {Schmeisser}, {Schmidt}, {Schneider}, {Schr{\"o}der},
  {Schumacher}, {Sclafani}, {Seckel}, {Seikh}, {Seo}, {Seunarine}, {Sevle
  Myhr}, {Shah}, {Shefali}, {Shimizu}, {Silva}, {Skrzypek}, {Smithers},
  {Snihur}, {Soedingrekso}, {S{\o}gaard}, {Soldin}, {Soldin}, {Sommani},
  {Spannfellner}, {Spiczak}, {Spiering}, {Stamatikos}, {Stanev},
  {Stezelberger}, {St{\"u}rwald}, {Stuttard}, {Sullivan}, {Taboada},
  {Ter-Antonyan}, {Terliuk}, {Thiesmeyer}, {Thompson}, {Thwaites}, {Tilav},
  {Tollefson}, {T{\"o}nnis}, {Toscano}, {Tosi}, {Trettin}, {Turcotte},
  {Twagirayezu}, {Unland Elorrieta}, {Upadhyay}, {Upshaw}, {Vaidyanathan},
  {Valtonen-Mattila}, {Vandenbroucke}, {van Eijndhoven}, {Vannerom}, {van
  Santen}, {Vara}, {Varsi}, {Veitch-Michaelis}, {Venugopal}, {Vereecken},
  {Verpoest}, {Veske}, {Vijai}, {Walck}, {Wang}, {Weaver}, {Weigel}, {Weindl},
  {Weldert}, {Wen}, {Wendt}, {Werthebach}, {Weyrauch}, {Whitehorn}, {Wiebusch},
  {Williams}, {Witthaus}, {Wolf}, {Wolf}, {Wrede}, {Xu}, {Yanez}, {Yildizci},
  {Yoshida}, {Young}, {Yu}, {Yuan}, {Zhang}, {Zhelnin}, {Zilberman}, \&
  {Zimmerman}}]{Abbasi24b}
{Abbasi}, R., {Ackermann}, M., {Adams}, J., {et~al.} 2024{\natexlab{b}}, arXiv
  e-prints, arXiv:2406.07601

\bibitem[{{Abbasi} {et~al.}(2021){Abbasi}, {Ackermann}, {Adams}, {Aguilar},
  {Ahlers}, {Ahrens}, {Alispach}, {Alves}, {Amin}, {An}, {Andeen}, {Anderson},
  {Anton}, {Arg{\"u}elles}, {Ashida}, {Axani}, {Bai}, {Balagopal}, {Barbano},
  {Barwick}, {Bastian}, {Basu}, {Baur}, {Bay}, {Beatty}, {Becker}, {Becker
  Tjus}, {Bellenghi}, {BenZvi}, {Berley}, {Bernardini}, {Besson}, {Binder},
  {Bindig}, {Blaufuss}, {Blot}, {Boddenberg}, {Bontempo}, {Borowka},
  {B{\"o}ser}, {Botner}, {B{\"o}ttcher}, {Bourbeau}, {Bradascio}, {Braun},
  {Bron}, {Brostean-Kaiser}, {Browne}, {Burgman}, {Burley}, {Busse}, {Campana},
  {Carnie-Bronca}, {Chen}, {Chirkin}, {Choi}, {Clark}, {Clark}, {Classen},
  {Coleman}, {Collin}, {Conrad}, {Coppin}, {Correa}, {Cowen}, {Cross},
  {Dappen}, {Dave}, {De Clercq}, {DeLaunay}, {Dembinski}, {Deoskar}, {De
  Ridder}, {Desai}, {Desiati}, {de Vries}, {de Wasseige}, {de With}, {DeYoung},
  {Dharani}, {Diaz}, {D{\'\i}az-V{\'e}lez}, {Dittmer}, {Dujmovic}, {Dunkman},
  {DuVernois}, {Dvorak}, {Ehrhardt}, {Eller}, {Engel}, {Erpenbeck}, {Evans},
  {Evenson}, {Fan}, {Fazely}, {Fiedlschuster}, {Fienberg}, {Filimonov},
  {Finley}, {Fischer}, {Fox}, {Franckowiak}, {Friedman}, {Fritz}, {F{\"u}rst},
  {Gaisser}, {Gallagher}, {Ganster}, {Garcia}, {Garrappa}, {Gerhardt},
  {Ghadimi}, {Glaser}, {Glauch}, {Gl{\"u}senkamp}, {Goldschmidt}, {Gonzalez},
  {Goswami}, {Grant}, {Gr{\'e}goire}, {Griswold}, {G{\"u}nd{\"u}z},
  {G{\"u}nther}, {Haack}, {Hallgren}, {Halliday}, {Halve}, {Halzen}, {Ha Minh},
  {Hanson}, {Hardin}, {Harnisch}, {Haungs}, {Hauser}, {Hebecker}, {Helbing},
  {Henningsen}, {Hettinger}, {Hickford}, {Hignight}, {Hill}, {Hill}, {Hoffman},
  {Hoffmann}, {Hoinka}, {Hokanson-Fasig}, {Hoshina}, {Huang}, {Huber}, {Huber},
  {Hultqvist}, {H{\"u}nnefeld}, {Hussain}, {In}, {Iovine}, {Ishihara},
  {Jansson}, {Japaridze}, {Jeong}, {Jones}, {Kang}, {Kang}, {Kang}, {Kappes},
  {Kappesser}, {Karg}, {Karl}, {Karle}, {Katz}, {Kauer}, {Kellermann},
  {Kelley}, {Kheirandish}, {Kin}, {Kintscher}, {Kiryluk}, {Klein}, {Koirala},
  {Kolanoski}, {Kontrimas}, {K{\"o}pke}, {Kopper}, {Kopper}, {Koskinen},
  {Koundal}, {Kovacevich}, {Kowalski}, {Kozynets}, {Kun}, {Kurahashi}, {Lad},
  {Lagunas Gualda}, {Lanfranchi}, {Larson}, {Lauber}, {Lazar}, {Lee},
  {Leonard}, {Leszczy{\'n}ska}, \& {Li}}]{NGC1068_steady_1}
{Abbasi}, R., {Ackermann}, M., {Adams}, J., {et~al.} 2021, \apjl, 920, L45

\bibitem[{{Abdollahi} {et~al.}(2022){Abdollahi}, {Acero}, {Baldini}, {Ballet},
  {Bastieri}, {Bellazzini}, {Berenji}, {Berretta}, {Bissaldi}, {Blandford},
  {Bloom}, {Bonino}, {Brill}, {Britto}, {Bruel}, {Burnett}, {Buson}, {Cameron},
  {Caputo}, {Caraveo}, {Castro}, {Chaty}, {Cheung}, {Chiaro}, {Cibrario},
  {Ciprini}, {Coronado-Bl{\'a}zquez}, {Crnogorcevic}, {Cutini}, {D'Ammando},
  {De Gaetano}, {Digel}, {Di Lalla}, {Dirirsa}, {Di Venere}, {Dom{\'\i}nguez},
  {Fallah Ramazani}, {Fegan}, {Ferrara}, {Fiori}, {Fleischhack}, {Franckowiak},
  {Fukazawa}, {Funk}, {Fusco}, {Galanti}, {Gammaldi}, {Gargano}, {Garrappa},
  {Gasparrini}, {Giacchino}, {Giglietto}, {Giordano}, {Giroletti}, {Glanzman},
  {Green}, {Grenier}, {Grondin}, {Guillemot}, {Guiriec}, {Gustafsson},
  {Harding}, {Hays}, {Hewitt}, {Horan}, {Hou}, {J{\'o}hannesson}, {Karwin},
  {Kayanoki}, {Kerr}, {Kuss}, {Landriu}, {Larsson}, {Latronico},
  {Lemoine-Goumard}, {Li}, {Liodakis}, {Longo}, {Loparco}, {Lott}, {Lubrano},
  {Maldera}, {Malyshev}, {Manfreda}, {Mart{\'\i}-Devesa}, {Mazziotta}, {Mereu},
  {Meyer}, {Michelson}, {Mirabal}, {Mitthumsiri}, {Mizuno}, {Moiseev},
  {Monzani}, {Morselli}, {Moskalenko}, {Negro}, {Nuss}, {Omodei}, {Orienti},
  {Orlando}, {Paneque}, {Pei}, {Perkins}, {Persic}, {Pesce-Rollins},
  {Petrosian}, {Pillera}, {Poon}, {Porter}, {Principe}, {Rain{\`o}}, {Rando},
  {Rani}, {Razzano}, {Razzaque}, {Reimer}, {Reimer}, {Reposeur},
  {S{\'a}nchez-Conde}, {Saz Parkinson}, {Scotton}, {Serini}, {Sgr{\`o}},
  {Siskind}, {Smith}, {Spandre}, {Spinelli}, {Sueoka}, {Suson}, {Tajima},
  {Tak}, {Thayer}, {Thompson}, {Torres}, {Troja}, {Valverde}, {Wood}, \&
  {Zaharijas}}]{FermiDR3}
{Abdollahi}, S., {Acero}, F., {Baldini}, L., {et~al.} 2022, \apjs, 260, 53

\bibitem[{{Acciari} {et~al.}(2019){Acciari}, {Ansoldi}, {Antonelli}, {Arbet
  Engels}, {Baack}, {Babi{\'c}}, {Banerjee}, {Barres de Almeida}, {Barrio},
  {Becerra Gonz{\'a}lez}, {Bednarek}, {Bellizzi}, {Bernardini}, {Berti},
  {Besenrieder}, {Bhattacharyya}, {Bigongiari}, {Biland}, {Blanch}, {Bonnoli},
  {Bo{\v{s}}njak}, {Busetto}, {Carosi}, {Ceribella}, {Chai}, {Chilingaryan},
  {Cikota}, {Colak}, {Colin}, {Colombo}, {Contreras}, {Cortina}, {Covino},
  {D'Elia}, {Da Vela}, {Dazzi}, {De Angelis}, {De Lotto}, {Delfino}, {Delgado},
  {Depaoli}, {Di Pierro}, {Di Venere}, {Do Souto Espi{\~n}eira}, {Dominis
  Prester}, {Donini}, {Dorner}, {Doro}, {Elsaesser}, {Fallah Ramazani},
  {Fattorini}, {Ferrara}, {Fidalgo}, {Foffano}, {Fonseca}, {Font}, {Fruck},
  {Fukami}, {Garc{\'\i}a L{\'o}pez}, {Garczarczyk}, {Gasparyan}, {Gaug},
  {Giglietto}, {Giordano}, {Godinovi{\'c}}, {Green}, {Guberman}, {Hadasch},
  {Hahn}, {Herrera}, {Hoang}, {Hrupec}, {H{\"u}tten}, {Inada}, {Inoue},
  {Ishio}, {Iwamura}, {Jouvin}, {Kerszberg}, {Kubo}, {Kushida}, {Lamastra},
  {Lelas}, {Leone}, {Lindfors}, {Lombardi}, {Longo}, {L{\'o}pez},
  {L{\'o}pez-Coto}, {L{\'o}pez-Oramas}, {Loporchio}, {Machado de Oliveira
  Fraga}, {Maggio}, {Majumdar}, {Makariev}, {Mallamaci}, {Maneva}, {Manganaro},
  {Mannheim}, {Maraschi}, {Mariotti}, {Mart{\'\i}nez}, {Mazin},
  {Mi{\'c}anovi{\'c}}, {Miceli}, {Minev}, {Miranda}, {Mirzoyan}, {Molina},
  {Moralejo}, {Morcuende}, {Moreno}, {Moretti}, {Munar-Adrover}, {Neustroev},
  {Nigro}, {Nilsson}, {Ninci}, {Nishijima}, {Noda}, {Nogu{\'e}s}, {Nozaki},
  {Paiano}, {Palacio}, {Palatiello}, {Paneque}, {Paoletti}, {Paredes},
  {Pe{\~n}il}, {Peresano}, {Persic}, {Prada Moroni}, {Prandini}, {Puljak},
  {Rhode}, {Rib{\'o}}, {Rico}, {Righi}, {Rugliancich}, {Saha}, {Sahakyan},
  {Saito}, {Sakurai}, {Satalecka}, {Schmidt}, {Schweizer}, {Sitarek},
  {{\v{S}}nidari{\'c}}, {Sobczynska}, {Somero}, {Stamerra}, {Strom}, {Strzys},
  {Suda}, {Suri{\'c}}, {Takahashi}, {Tavecchio}, {Temnikov}, {Terzi{\'c}},
  {Teshima}, {Torres-Alb{\`a}}, {Tosti}, {Vagelli}, {van Scherpenberg},
  {Vanzo}, {Vazquez Acosta}, {Vigorito}, {Vitale}, {Vovk}, {Will}, {Zari{\'c}},
  {MAGIC Collaboration}, {Fiore}, {Feruglio}, \& {Rephaeli}}]{Acciari19}
{Acciari}, V.~A., {Ansoldi}, S., {Antonelli}, L.~A., {et~al.} 2019, \apj, 883,
  135

\bibitem[{{Ackermann} {et~al.}(2015){Ackermann}, {Ajello}, {Albert}, {Atwood},
  {Baldini}, {Ballet}, {Barbiellini}, {Bastieri}, {Bechtol}, {Bellazzini},
  {Bissaldi}, {Blandford}, {Bloom}, {Bottacini}, {Brandt}, {Bregeon}, {Bruel},
  {Buehler}, {Buson}, {Caliandro}, {Cameron}, {Caragiulo}, {Caraveo},
  {Cavazzuti}, {Cecchi}, {Charles}, {Chekhtman}, {Chiang}, {Chiaro}, {Ciprini},
  {Claus}, {Cohen-Tanugi}, {Conrad}, {Cuoco}, {Cutini}, {D'Ammando}, {de
  Angelis}, {de Palma}, {Dermer}, {Digel}, {Silva}, {Drell}, {Favuzzi},
  {Ferrara}, {Focke}, {Franckowiak}, {Fukazawa}, {Funk}, {Fusco}, {Gargano},
  {Gasparrini}, {Germani}, {Giglietto}, {Giommi}, {Giordano}, {Giroletti},
  {Godfrey}, {Gomez-Vargas}, {Grenier}, {Guiriec}, {Gustafsson}, {Hadasch},
  {Hayashi}, {Hays}, {Hewitt}, {Ippoliti}, {Jogler}, {J{\'o}hannesson},
  {Johnson}, {Johnson}, {Kamae}, {Kataoka}, {Kn{\"o}dlseder}, {Kuss},
  {Larsson}, {Latronico}, {Li}, {Li}, {Longo}, {Loparco}, {Lott}, {Lovellette},
  {Lubrano}, {Madejski}, {Manfreda}, {Massaro}, {Mayer}, {Mazziotta},
  {McEnery}, {Michelson}, {Mitthumsiri}, {Mizuno}, {Moiseev}, {Monzani},
  {Morselli}, {Moskalenko}, {Murgia}, {Nemmen}, {Nuss}, {Ohsugi}, {Omodei},
  {Orlando}, {Ormes}, {Paneque}, {Panetta}, {Perkins}, {Pesce-Rollins},
  {Piron}, {Pivato}, {Porter}, {Rain{\`o}}, {Rando}, {Razzano}, {Razzaque},
  {Reimer}, {Reimer}, {Reposeur}, {Ritz}, {Romani}, {S{\'a}nchez-Conde},
  {Schaal}, {Schulz}, {Sgr{\`o}}, {Siskind}, {Spandre}, {Spinelli}, {Strong},
  {Suson}, {Takahashi}, {Thayer}, {Thayer}, {Tibaldo}, {Tinivella}, {Torres},
  {Tosti}, {Troja}, {Uchiyama}, {Vianello}, {Werner}, {Winer}, {Wood}, {Wood},
  {Zaharijas}, \& {Zimmer}}]{Ackermann15}
{Ackermann}, M., {Ajello}, M., {Albert}, A., {et~al.} 2015, \apj, 799, 86

\bibitem[{Aharonian {et~al.}(2008)Aharonian, Khangulyan, \&
  Costamante}]{Aharonian_2008}
Aharonian, F., Khangulyan, D., \& Costamante, L. 2008, Mon. Not. Roy. Astron.
  Soc., 387, 1206

\bibitem[{{Aharonian}(2004)}]{Aharonian04}
{Aharonian}, F.~A. 2004, {Very high energy cosmic gamma radiation : a crucial
  window on the extreme Universe}

\bibitem[{{Ajello} {et~al.}(2021){Ajello}, {Baldini}, {Ballet}, {Barbiellini},
  {Bastieri}, {Bellazzini}, {Berretta}, {Bissaldi}, {Blandford}, {Bloom},
  {Bonino}, {Bruel}, {Buson}, {Cameron}, {Caprioli}, {Caputo}, {Cavazzuti},
  {Chartas}, {Chen}, {Cheung}, {Chiaro}, {Costantin}, {Cutini}, {D'Ammando},
  {de la Torre Luque}, {de Palma}, {Desai}, {Diesing}, {Di Lalla}, {Dirirsa},
  {Di Venere}, {Dom{\'\i}nguez}, {Fegan}, {Franckowiak}, {Fukazawa}, {Funk},
  {Fusco}, {Gargano}, {Gasparrini}, {Giglietto}, {Giordano}, {Giroletti},
  {Green}, {Grenier}, {Guiriec}, {Hartmann}, {Horan}, {J{\'o}hannesson},
  {Karwin}, {Kerr}, {Kova{\v{c}}evi{\'c}}, {Kuss}, {Larsson}, {Latronico},
  {Lemoine-Goumard}, {Li}, {Liodakis}, {Longo}, {Loparco}, {Lovellette},
  {Lubrano}, {Maldera}, {Manfreda}, {Marchesi}, {Marcotulli},
  {Mart{\'\i}-Devesa}, {Mazziotta}, {Mereu}, {Michelson}, {Mizuno}, {Monzani},
  {Morselli}, {Moskalenko}, {Negro}, {Omodei}, {Orienti}, {Orlando}, {Paliya},
  {Paneque}, {Pei}, {Persic}, {Pesce-Rollins}, {Porter}, {Principe}, {Racusin},
  {Rain{\`o}}, {Rando}, {Rani}, {Razzano}, {Reimer}, {Reimer}, {Saz Parkinson},
  {Serini}, {Sgr{\`o}}, {Siskind}, {Spandre}, {Spinelli}, {Suson}, {Tak},
  {Torres}, {Troja}, {Wood}, {Zaharijas}, \& {Zrake}}]{Ajello_UFO}
{Ajello}, M., {Baldini}, L., {Ballet}, J., {et~al.} 2021, \apj, 921, 144

\bibitem[{{Ajello} {et~al.}(2023){Ajello}, {Murase}, \& {McDaniel}}]{Ajello23}
{Ajello}, M., {Murase}, K., \& {McDaniel}, A. 2023, \apjl, 954, L49

\bibitem[{{Aleksi{\'c}} {et~al.}(2016){Aleksi{\'c}}, {Ansoldi}, {Antonelli},
  {Antoranz}, {Babic}, {Bangale}, {Barcel{\'o}}, {Barrio}, {Becerra
  Gonz{\'a}lez}, {Bednarek}, {Bernardini}, {Biasuzzi}, {Biland}, {Bitossi},
  {Blanch}, {Bonnefoy}, {Bonnoli}, {Borracci}, {Bretz}, {Carmona}, {Carosi},
  {Cecchi}, {Colin}, {Colombo}, {Contreras}, {Corti}, {Cortina}, {Covino}, {Da
  Vela}, {Dazzi}, {De Angelis}, {De Caneva}, {De Lotto}, {de O{\~n}a Wilhelmi},
  {Delgado Mendez}, {Dettlaff}, {Dominis Prester}, {Dorner}, {Doro}, {Einecke},
  {Eisenacher}, {Elsaesser}, {Fidalgo}, {Fink}, {Fonseca}, {Font}, {Frantzen},
  {Fruck}, {Galindo}, {Garc{\'\i}a L{\'o}pez}, {Garczarczyk}, {Garrido
  Terrats}, {Gaug}, {Giavitto}, {Godinovi{\'c}}, {Gonz{\'a}lez Mu{\~n}oz},
  {Gozzini}, {Haberer}, {Hadasch}, {Hanabata}, {Hayashida}, {Herrera},
  {Hildebrand}, {Hose}, {Hrupec}, {Idec}, {Illa}, {Kadenius}, {Kellermann},
  {Knoetig}, {Kodani}, {Konno}, {Krause}, {Kubo}, {Kushida}, {La Barbera},
  {Lelas}, {Lemus}, {Lewandowska}, {Lindfors}, {Lombardi}, {Longo},
  {L{\'o}pez}, {L{\'o}pez-Coto}, {L{\'o}pez-Oramas}, {Lorca}, {Lorenz},
  {Lozano}, {Makariev}, {Mallot}, {Maneva}, {Mankuzhiyil}, {Mannheim},
  {Maraschi}, {Marcote}, {Mariotti}, {Mart{\'\i}nez}, {Mazin}, {Menzel},
  {Miranda}, {Mirzoyan}, {Moralejo}, {Munar-Adrover}, {Nakajima}, {Negrello},
  {Neustroev}, {Niedzwiecki}, {Nilsson}, {Nishijima}, {Noda}, {Orito},
  {Overkemping}, {Paiano}, {Palatiello}, {Paneque}, {Paoletti}, {Paredes},
  {Paredes-Fortuny}, {Persic}, {Poutanen}, {Prada Moroni}, {Prandini},
  {Puljak}, {Reinthal}, {Rhode}, {Rib{\'o}}, {Rico}, {Rodriguez Garcia},
  {R{\"u}gamer}, {Saito}, {Saito}, {Satalecka}, {Scalzotto}, {Scapin},
  {Schultz}, {Schlammer}, {Schmidl}, {Schweizer}, {Shore}, {Sillanp{\"a}{\"a}},
  {Sitarek}, {Snidaric}, {Sobczynska}, {Spanier}, {Stamerra}, {Steinbring},
  {Storz}, {Strzys}, {Takalo}, {Takami}, {Tavecchio}, {Tejedor}, {Temnikov},
  {Terzi{\'c}}, {Tescaro}, {Teshima}, {Thaele}, {Tibolla}, {Torres}, {Toyama},
  {Treves}, {Vogler}, {Wetteskind}, {Will}, \& {Zanin}}]{Aleksic16}
{Aleksi{\'c}}, J., {Ansoldi}, S., {Antonelli}, L.~A., {et~al.} 2016,
  Astroparticle Physics, 72, 76

\bibitem[{Aleksić {et~al.}(2016)Aleksić, Ansoldi, Antonelli, Antoranz, Babic,
  Bangale, Barceló, Barrio, {Becerra González}, Bednarek, Bernardini,
  Biasuzzi, Biland, Bitossi, Blanch, Bonnefoy, Bonnoli, Borracci, Bretz,
  Carmona, Carosi, Cecchi, Colin, Colombo, Contreras, Corti, Cortina, Covino,
  {Da Vela}, Dazzi, DeAngelis, {De Caneva}, {De Lotto}, {de Oña Wilhelmi},
  {Delgado Mendez}, Dettlaff, {Dominis Prester}, Dorner, Doro, Einecke,
  Eisenacher, Elsaesser, Fidalgo, Fink, Fonseca, Font, Frantzen, Fruck,
  Galindo, {García López}, Garczarczyk, {Garrido Terrats}, Gaug, Giavitto,
  Godinović, {González Muñoz}, Gozzini, Haberer, Hadasch, Hanabata,
  Hayashida, Herrera, Hildebrand, Hose, Hrupec, Idec, Illa, Kadenius,
  Kellermann, Knoetig, Kodani, Konno, Krause, Kubo, Kushida, {La Barbera},
  Lelas, Lemus, Lewandowska, Lindfors, Lombardi, Longo, López, López-Coto,
  López-Oramas, Lorca, Lorenz, Lozano, Makariev, Mallot, Maneva, Mankuzhiyil,
  Mannheim, Maraschi, Marcote, Mariotti, Martínez, Mazin, Menzel, Miranda,
  Mirzoyan, Moralejo, Munar-Adrover, Nakajima, Negrello, Neustroev,
  Niedzwiecki, Nilsson, Nishijima, Noda, Orito, Overkemping, Paiano,
  Palatiello, Paneque, Paoletti, Paredes, Paredes-Fortuny, Persic, Poutanen,
  {Prada Moroni}, Prandini, Puljak, Reinthal, Rhode, Ribó, Rico, {Rodriguez
  Garcia}, Rügamer, Saito, Saito, Satalecka, Scalzotto, Scapin, Schultz,
  Schlammer, Schmidl, Schweizer, Sillanpää, Sitarek, Snidaric, Sobczynska,
  Spanier, Stamerra, Steinbring, Storz, Strzys, Takalo, Takami, Tavecchio,
  Tejedor, Temnikov, Terzić, Tescaro, Teshima, Thaele, Tibolla, Torres,
  Toyama, Treves, Vogler, Wetteskind, Will, \& Zanin}]{Aleksic16A}
Aleksić, J., Ansoldi, S., Antonelli, L., {et~al.} 2016, Astroparticle Physics,
  72, 61

\bibitem[{{Ambrosone}(2024)}]{Ambrosone24}
{Ambrosone}, A. 2024, \jcap, 2024, 075

\bibitem[{{Ballet} {et~al.}(2023)}]{fermilatdr4}
{Ballet}, J. {et~al.} 2023, arXiv e-prints, arXiv:2307.12546

\bibitem[{{Bechtol} {et~al.}(2017){Bechtol}, {Ahlers}, {Di Mauro}, {Ajello}, \&
  {Vandenbroucke}}]{Bechtol17}
{Bechtol}, K., {Ahlers}, M., {Di Mauro}, M., {Ajello}, M., \& {Vandenbroucke},
  J. 2017, \apj, 836, 47

\bibitem[{{Bentz} {et~al.}(2006){Bentz}, {Denney}, {Cackett}, {Dietrich},
  {Fogel}, {Ghosh}, {Horne}, {Kuehn}, {Minezaki}, {Onken}, {Peterson}, {Pogge},
  {Pronik}, {Richstone}, {Sergeev}, {Vestergaard}, {Walker}, \&
  {Yoshii}}]{Bentz_2006}
{Bentz}, M.~C., {Denney}, K.~D., {Cackett}, E.~M., {et~al.} 2006, \apj, 651,
  775

\bibitem[{{Berezinsky}(1977)}]{Berezinsky_1977}
{Berezinsky}, V.~S. 1977, in Proceedings of the International Conference
  Neutrino ’77

\bibitem[{Berezinsky {et~al.}(1990)Berezinsky, Bulanov, Dogiel, \&
  Ptuskin}]{Berezinsky:1990qxi}
Berezinsky, V.~S., Bulanov, S.~V., Dogiel, V.~A., \& Ptuskin, V.~S. 1990,
  {Astrophysics of cosmic rays}, ed. V.~L. Ginzburg

\bibitem[{{Berezinsky} \& {Ginzburg}(1981)}]{Berezinsky_1981}
{Berezinsky}, V.~S. \& {Ginzburg}, V.~L. 1981, in International Cosmic Ray
  Conference, Vol.~1, International Cosmic Ray Conference, 238

\bibitem[{{Condorelli} \& {Petrera}(2025)}]{Condorelli25}
{Condorelli}, A. \& {Petrera}, S. 2025, Astroparticle Physics, 165, 103047

\bibitem[{{Dave} {et~al.}(2024){Dave}, {Taboada}, \& {IceCube
  Collaboration}}]{NGC1068_steady_2}
{Dave}, P., {Taboada}, I., \& {IceCube Collaboration}. 2024, in 38th
  International Cosmic Ray Conference, 973

\bibitem[{{Eichler}(1979)}]{Eichler_1979}
{Eichler}, D. 1979, \apj, 232, 106

\bibitem[{{Eichmann} {et~al.}(2022){Eichmann}, {Oikonomou}, {Salvatore},
  {Dettmar}, \& {Tjus}}]{Eichmann20}
{Eichmann}, B., {Oikonomou}, F., {Salvatore}, S., {Dettmar}, R.-J., \& {Tjus},
  J.~B. 2022, \apj, 939, 43

\bibitem[{{Fang} {et~al.}(2022){Fang}, {Gallagher}, \& {Halzen}}]{Fang_2022}
{Fang}, K., {Gallagher}, J.~S., \& {Halzen}, F. 2022, \apj, 933, 190

\bibitem[{{Fiorillo} {et~al.}(2024{\natexlab{a}}){Fiorillo}, {Comisso},
  {Peretti}, {Petropoulou}, \& {Sironi}}]{Fiorillo24b}
{Fiorillo}, D. F.~G., {Comisso}, L., {Peretti}, E., {Petropoulou}, M., \&
  {Sironi}, L. 2024{\natexlab{a}}, \apj, 974, 75

\bibitem[{{Fiorillo} {et~al.}(2025){Fiorillo}, {Comisso}, {Peretti},
  {Petropoulou}, \& {Sironi}}]{Fiorillo25}
{Fiorillo}, D. F.~G., {Comisso}, L., {Peretti}, E., {Petropoulou}, M., \&
  {Sironi}, L. 2025, arXiv e-prints, arXiv:2504.06336

\bibitem[{{Fiorillo} {et~al.}(2024{\natexlab{b}}){Fiorillo}, {Petropoulou},
  {Comisso}, {Peretti}, \& {Sironi}}]{Fiorillo24a}
{Fiorillo}, D. F.~G., {Petropoulou}, M., {Comisso}, L., {Peretti}, E., \&
  {Sironi}, L. 2024{\natexlab{b}}, \apjl, 961, L14

\bibitem[{{Fomin} {et~al.}(1994){Fomin}, {Stepanian}, {Lamb}, {Lewis}, {Punch},
  \& {Weekes}}]{Fomin94}
{Fomin}, V.~P., {Stepanian}, A.~A., {Lamb}, R.~C., {et~al.} 1994, Astroparticle
  Physics, 2, 137

\bibitem[{{Franceschini} \& {Rodighiero}(2017)}]{Franceschini_EBL}
{Franceschini}, A. \& {Rodighiero}, G. 2017, \aap, 603, A34

\bibitem[{{Fruck} {et~al.}(2022){Fruck}, {Gaug}, {Hahn}, {Acciari},
  {Besenrieder}, {Dominis Prester}, {Dorner}, {Fink}, {Font},
  {Mi{\'c}anovi{\'c}}, {Mirzoyan}, {M{\"u}ller}, {Pavleti{\'c}},
  {Schmuckermaier}, \& {Will}}]{LIDAR_1}
{Fruck}, C., {Gaug}, M., {Hahn}, A., {et~al.} 2022, \mnras, 515, 4520

\bibitem[{{Gianolli} {et~al.}(2023){Gianolli}, {Kim}, {Bianchi},
  {Ag{\'\i}s-Gonz{\'a}lez}, {Madejski}, {Marin}, {Marinucci}, {Matt}, {Middei},
  {Petrucci}, {Soffitta}, {Tagliacozzo}, {Tombesi}, {Ursini}, {Barnouin}, {De
  Rosa}, {Di Gesu}, {Ingram}, {Loktev}, {Panagiotou}, {Podgorny}, {Poutanen},
  {Puccetti}, {Ratheesh}, {Veledina}, {Zhang}, {Agudo}, {Antonelli},
  {Bachetti}, {Baldini}, {Baumgartner}, {Bellazzini}, {Bongiorno}, {Bonino},
  {Brez}, {Bucciantini}, {Capitanio}, {Castellano}, {Cavazzuti}, {Chen},
  {Ciprini}, {Costa}, {Del Monte}, {Di Lalla}, {Di Marco}, {Donnarumma},
  {Doroshenko}, {Dov{\v{c}}iak}, {Ehlert}, {Enoto}, {Evangelista}, {Fabiani},
  {Ferrazzoli}, {Garc{\'\i}a}, {Gunji}, {Heyl}, {Iwakiri}, {Jorstad}, {Kaaret},
  {Karas}, {Kislat}, {Kitaguchi}, {Kolodziejczak}, {Krawczynski}, {La Monaca},
  {Latronico}, {Liodakis}, {Maldera}, {Manfreda}, {Marscher}, {Marshall},
  {Massaro}, {Mitsuishi}, {Mizuno}, {Muleri}, {Negro}, {Ng}, {O'Dell},
  {Omodei}, {Oppedisano}, {Papitto}, {Pavlov}, {Peirson}, {Perri},
  {Pesce-Rollins}, {Pilia}, {Possenti}, {Ramsey}, {Rankin}, {Roberts},
  {Romani}, {Sgr{\`o}}, {Slane}, {Spandre}, {Swartz}, {Tamagawa}, {Tavecchio},
  {Taverna}, {Tawara}, {Tennant}, {Thomas}, {Trois}, {Tsygankov}, {Turolla},
  {Vink}, {Weisskopf}, {Wu}, {Xie}, \& {Zane}}]{Gianolli_2023}
{Gianolli}, V.~E., {Kim}, D.~E., {Bianchi}, S., {et~al.} 2023, \mnras, 523,
  4468

\bibitem[{{IceCube Collaboration}(2013)}]{IceCube2013_Science}
{IceCube Collaboration}. 2013, Science, 342, 1242856

\bibitem[{{IceCube Collaboration} {et~al.}(2022){IceCube Collaboration},
  {Abbasi}, {Ackermann}, {Adams}, {Aguilar}, {Ahlers}, {Ahrens}, {Alameddine},
  {Alispach}, {Alves}, {Amin}, {Andeen}, {Anderson}, {Anton}, {Arg{\"u}elles},
  {Ashida}, {Axani}, {Bai}, {Balagopal}, {Barbano}, {Barwick}, {Bastian},
  {Basu}, {Baur}, {Bay}, {Beatty}, {Becker}, {Becker Tjus}, {Bellenghi},
  {Benzvi}, {Berley}, {Bernardini}, {Besson}, {Binder}, {Bindig}, {Blaufuss},
  {Blot}, {Boddenberg}, {Bontempo}, {Borowka}, {B{\"o}ser}, {Botner},
  {B{\"o}ttcher}, {Bourbeau}, {Bradascio}, {Braun}, {Brinson}, {Bron},
  {Brostean-Kaiser}, {Browne}, {Burgman}, {Burley}, {Busse}, {Campana},
  {Carnie-Bronca}, {Chen}, {Chen}, {Chirkin}, {Choi}, {Clark}, {Clark},
  {Classen}, {Coleman}, {Collin}, {Conrad}, {Coppin}, {Correa}, {Cowen},
  {Cross}, {Dappen}, {Dave}, {de Clercq}, {Delaunay}, {Delgado L{\'o}pez},
  {Dembinski}, {Deoskar}, {Desai}, {Desiati}, {de Vries}, {de Wasseige}, {de
  With}, {Deyoung}, {Diaz}, {D{\'\i}az-V{\'e}lez}, {Dittmer}, {Dujmovic},
  {Dunkman}, {Duvernois}, {Dvorak}, {Ehrhardt}, {Eller}, {Engel}, {Erpenbeck},
  {Evans}, {Evenson}, {Fan}, {Fazely}, {Fedynitch}, {Feigl}, {Fiedlschuster},
  {Fienberg}, {Filimonov}, {Finley}, {Fischer}, {Fox}, {Franckowiak},
  {Friedman}, {Fritz}, {F{\"u}rst}, {Gaisser}, {Gallagher}, {Ganster},
  {Garcia}, {Garrappa}, {Gerhardt}, {Ghadimi}, {Glaser}, {Glauch},
  {Gl{\"u}senkamp}, {Goldschmidt}, {Gonzalez}, {Goswami}, {Grant},
  {Gr{\'e}goire}, {Griswold}, {G{\"u}nther}, {Gutjahr}, {Haack}, {Hallgren},
  {Halliday}, {Halve}, {Halzen}, {Hanson}, {Hardin}, {Harnisch}, {Haungs},
  {Hebecker}, {Helbing}, {Henningsen}, {Hettinger}, {Hickford}, {Hignight},
  {Hill}, {Hill}, {Hoffman}, {Hoffmann}, {Hokanson-Fasig}, {Hoshina}, {Huang},
  {Huber}, {Huber}, {Hultqvist}, {H{\"u}nnefeld}, {Hussain}, {Hymon}, {in},
  {Iovine}, {Ishihara}, {Jansson}, {Japaridze}, {Jeong}, {Jin}, {Jones},
  {Kang}, {Kang}, {Kang}, {Kappes}, {Kappesser}, {Kardum}, {Karg}, {Karl},
  {Karle}, {Katz}, {Kauer}, {Kellermann}, {Kelley}, {Kheirandish}, {Kin},
  {Kintscher}, {Kiryluk}, {Klein}, {Koirala}, {Kolanoski}, {Kontrimas},
  {K{\"o}pke}, {Kopper}, {Kopper}, {Koskinen}, {Koundal}, {Kovacevich},
  {Kowalski}, {Kozynets}, {Kun}, {Kurahashi}, {Lad}, {Lagunas Gualda},
  {Lanfranchi}, {Larson}, {Lauber}, {Lazar}, {Lee}, {Leonard},
  {Leszczy{\'n}ska}, {Li}, {Lincetto}, {Liu}, {Liubarska}, {Lohfink}, {Lozano
  Mariscal}, {Lu}, {Lucarelli}, {Ludwig}, {Luszczak}, {Lyu}, {Ma}, {Madsen},
  {Mahn}, {Makino}, {Mancina}, {Mari{\c{s}}}, {Martinez-Soler}, {Maruyama},
  {Mase}, {McElroy}, {McNally}, {Mead}, {Meagher}, {Mechbal}, {Medina},
  {Meier}, {Meighen-Berger}, {Micallef}, {Mockler}, {Montaruli}, {Moore},
  {Morse}, {Moulai}, {Naab}, {Nagai}, {Nahnhauer}, {Naumann}, {Necker},
  {Nguyen}, {Niederhausen}, {Nisa}, {Nowicki}, {Nygren}, {Obertack},
  {Pollmann}, {Oehler}, {Oeyen}, {Olivas}, {O'Sullivan}, {Pandya}, {Pankova},
  {Park}, {Parker}, {Paudel}, {Paul}, {P{\'e}rez de Los Heros}, {Peters},
  {Peterson}, {Philippen}, {Pieper}, {Pittermann}, {Pizzuto}, {Plum},
  {Popovych}, {Porcelli}, {Prado Rodriguez}, {Price}, {Pries}, {Przybylski},
  {Rack-Helleis}, {Raissi}, {Rameez}, {Rawlins}, {Rea}, {Rehman},
  {Reichherzer}, {Reimann}, {Renzi}, {Resconi}, {Reusch}, {Rhode}, {Richman},
  {Riedel}, {Roberts}, {Robertson}, {Roellinghoff}, {Rongen}, {Rott}, {Ruhe},
  {Ryckbosch}, {Rysewyk Cantu}, {Safa}, {Saffer}, {Sanchez Herrera},
  {Sandrock}, {Sandroos}, {Santander}, {Sarkar}, {Sarkar}, {Satalecka},
  {Schaufel}, {Schieler}, {Schindler}, {Schmidt}, {Schneider}, {Schneider},
  {Schr{\"o}der}, {Schumacher}, {Schwefer}, {Sclafani}, {Seckel}, {Seunarine},
  {Sharma}, {Shefali}, {Silva}, {Skrzypek}, {Smithers}, {Snihur},
  {Soedingrekso}, {Soldin}, {Spannfellner}, {Spiczak}, {Spiering},
  {Stachurska}, {Stamatikos}, {Stanev}, {Stein}, {Stettner}, {Steuer},
  {Stezelberger}, {Stokstad}, {St{\"u}rwald}, {Stuttard}, {Sullivan},
  {Taboada}, {Ter-Antonyan}, {Tilav}, {Tischbein}, {Tollefson}, {T{\"o}nnis},
  {Toscano}, {Tosi}, {Trettin}, {Tselengidou}, {Tung}, {Turcati}, {Turcotte},
  {Turley}, {Twagirayezu}, {Ty}, {Unland Elorrieta}, {Valtonen-Mattila},
  {Vandenbroucke}, {van Eijndhoven}, {Vannerom}, {van Santen}, {Verpoest},
  {Walck}, {Watson}, {Weaver}, {Weigel}, {Weindl}, {Weiss}, {Weldert}, {Wendt},
  {Werthebach}, {Weyrauch}, {Whitehorn}, {Wiebusch}, {Williams}, {Wolf},
  {Woschnagg}, {Wrede}, {Wulff}, {Xu}, {Yanez}, {Yoshida}, {Yu}, {Yuan},
  {Zhangan}, \& {Zhelnin}}]{Abbasi22}
{IceCube Collaboration}, {Abbasi}, R., {Ackermann}, M., {et~al.} 2022, Science,
  378, 538

\bibitem[{{Inoue} {et~al.}(2022){Inoue}, {Cerruti}, {Murase}, \&
  {Liu}}]{Inoue2022}
{Inoue}, S., {Cerruti}, M., {Murase}, K., \& {Liu}, R.-Y. 2022, arXiv e-prints,
  arXiv:2207.02097

\bibitem[{{Inoue} {et~al.}(2020){Inoue}, {Khangulyan}, \& {Doi}}]{Inoue20}
{Inoue}, Y., {Khangulyan}, D., \& {Doi}, A. 2020, \apjl, 891, L33

\bibitem[{{Inoue} {et~al.}(2019){Inoue}, {Khangulyan}, {Inoue}, \&
  {Doi}}]{Inoue19}
{Inoue}, Y., {Khangulyan}, D., {Inoue}, S., \& {Doi}, A. 2019, \apj, 880, 40

\bibitem[{Kelner \& Aharonian(2008)}]{Kelner_pg}
Kelner, S.~R. \& Aharonian, F.~A. 2008, \prd, 78

\bibitem[{{Kelner} {et~al.}(2006){Kelner}, {Aharonian}, \&
  {Bugayov}}]{Kelner_pp}
{Kelner}, S.~R., {Aharonian}, F.~A., \& {Bugayov}, V.~V. 2006, \prd, 74, 034018

\bibitem[{Kheirandish {et~al.}(2021)Kheirandish, Murase, \&
  Kimura}]{Kheirandish21}
Kheirandish, A., Murase, K., \& Kimura, S.~S. 2021, \apj, 922, 45

\bibitem[{{Lafferty} \& {Wyatt}(1995)}]{Lafferty1995}
{Lafferty}, G.~D. \& {Wyatt}, T.~R. 1995, Nuclear Instruments and Methods in
  Physics Research A, 355, 541

\bibitem[{{Lemoine} \& {Rieger}(2025)}]{Lemoine:2024roa}
{Lemoine}, M. \& {Rieger}, F. 2025, \aap, 697, A124

\bibitem[{{Li} \& {Ma}(1983)}]{LiMa83}
{Li}, T.~P. \& {Ma}, Y.~Q. 1983, \apj, 272, 317

\bibitem[{{Lyu} \& {Rieke}(2021)}]{Lyu21_Torus}
{Lyu}, J. \& {Rieke}, G.~H. 2021, \apj, 912, 126

\bibitem[{{Marconi} {et~al.}(2004){Marconi}, {Risaliti}, {Gilli}, {Hunt},
  {Maiolino}, \& {Salvati}}]{Marconi_2004}
{Marconi}, A., {Risaliti}, G., {Gilli}, R., {et~al.} 2004, \mnras, 351, 169

\bibitem[{{Mbarek} {et~al.}(2024){Mbarek}, {Philippov}, {Chernoglazov},
  {Levinson}, \& {Mushotzky}}]{Mbarek24}
{Mbarek}, R., {Philippov}, A., {Chernoglazov}, A., {Levinson}, A., \&
  {Mushotzky}, R. 2024, \prd, 109, L101306

\bibitem[{{Mullaney} {et~al.}(2011){Mullaney}, {Alexander}, {Goulding}, \&
  {Hickox}}]{Mullaney_2011}
{Mullaney}, J.~R., {Alexander}, D.~M., {Goulding}, A.~D., \& {Hickox}, R.~C.
  2011, \mnras, 414, 1082

\bibitem[{{Murase}(2022)}]{Murase22}
{Murase}, K. 2022, \apjl, 941, L17

\bibitem[{{Murase} {et~al.}(2013){Murase}, {Ahlers}, \& {Lacki}}]{Murase2013}
{Murase}, K., {Ahlers}, M., \& {Lacki}, B.~C. 2013, \prd, 88, 121301

\bibitem[{{Murase} {et~al.}(2016){Murase}, {Guetta}, \& {Ahlers}}]{Murase2016}
{Murase}, K., {Guetta}, D., \& {Ahlers}, M. 2016, \prl, 116, 071101

\bibitem[{{Murase} {et~al.}(2024){Murase}, {Karwin}, {Kimura}, {Ajello}, \&
  {Buson}}]{Murase24}
{Murase}, K., {Karwin}, C.~M., {Kimura}, S.~S., {Ajello}, M., \& {Buson}, S.
  2024, \apjl, 961, L34

\bibitem[{{Murase} {et~al.}(2020){Murase}, {Kimura}, \&
  {M{\'e}sz{\'a}ros}}]{Murase20}
{Murase}, K., {Kimura}, S.~S., \& {M{\'e}sz{\'a}ros}, P. 2020, \prl, 125,
  011101

\bibitem[{{Neronov} {et~al.}(2024){Neronov}, {Savchenko}, \&
  {Semikoz}}]{Neronov24}
{Neronov}, A., {Savchenko}, D., \& {Semikoz}, D.~V. 2024, \prl, 132, 101002

\bibitem[{{Osterbrock} \& {Koski}(1976)}]{Osterbrock76}
{Osterbrock}, D.~E. \& {Koski}, A.~T. 1976, \mnras, 176, 61P

\bibitem[{{Padovani} {et~al.}(2024{\natexlab{a}}){Padovani}, {Gilli},
  {Resconi}, {Bellenghi}, \& {Henningsen}}]{Padovani_24_Seyfert}
{Padovani}, P., {Gilli}, R., {Resconi}, E., {Bellenghi}, C., \& {Henningsen},
  F. 2024{\natexlab{a}}, \aap, 684, L21

\bibitem[{{Padovani} {et~al.}(2024{\natexlab{b}}){Padovani}, {Resconi},
  {Ajello}, {Bellenghi}, {Bianchi}, {Blasi}, {Huang}, {Gabici}, {G{\'a}mez
  Rosas}, {Niederhausen}, {Peretti}, {Eichmann}, {Guetta}, {Lamastra}, \&
  {Shimizu}}]{Padovani24}
{Padovani}, P., {Resconi}, E., {Ajello}, M., {et~al.} 2024{\natexlab{b}},
  Nature Astronomy, 8, 1077

\bibitem[{{Peretti} {et~al.}(2020){Peretti}, {Blasi}, {Aharonian}, {Morlino},
  \& {Cristofari}}]{Peretti20}
{Peretti}, E., {Blasi}, P., {Aharonian}, F., {Morlino}, G., \& {Cristofari}, P.
  2020, \mnras, 493, 5880

\bibitem[{{Peretti} {et~al.}(2023){Peretti}, {Lamastra}, {Saturni}, {Ahlers},
  {Blasi}, {Morlino}, \& {Cristofari}}]{Peretti_UFO}
{Peretti}, E., {Lamastra}, A., {Saturni}, F.~G., {et~al.} 2023, \mnras, 526,
  181

\bibitem[{{Peretti} {et~al.}(2025){Peretti}, {Peron}, {Tombesi}, {Lamastra},
  {Saturni}, {Cerruti}, \& {Ahlers}}]{Peretti4151}
{Peretti}, E., {Peron}, G., {Tombesi}, F., {et~al.} 2025, \jcap, 2025, 013

\bibitem[{{Rolke} {et~al.}(2005){Rolke}, {L{\'o}pez}, \& {Conrad}}]{Rolke05}
{Rolke}, W.~A., {L{\'o}pez}, A.~M., \& {Conrad}, J. 2005, Nuclear Instruments
  and Methods in Physics Research A, 551, 493

\bibitem[{{Roulet} \& {Vissani}(2021)}]{Roulet21}
{Roulet}, E. \& {Vissani}, F. 2021, \jcap, 2021, 050

\bibitem[{{Schmuckermaier} {et~al.}(2023){Schmuckermaier}, {Gaug}, {Fruck},
  {Moralejo}, {Hahn}, {Dominis Prester}, {Dorner}, {Font}, {Mi{\'c}anovi{\'c}},
  {Mirzoyan}, {Paneque}, {Pavleti{\'c}}, {Sitarek}, \& {Will}}]{LIDAR_2}
{Schmuckermaier}, F., {Gaug}, M., {Fruck}, C., {et~al.} 2023, \aap, 673, A2

\bibitem[{{Silberberg} \& {Shapiro}(1979)}]{Silberberg_1979}
{Silberberg}, R. \& {Shapiro}, M.~M. 1979, in International Cosmic Ray
  Conference, Vol.~10, International Cosmic Ray Conference, 357

\bibitem[{{Stecker} {et~al.}(1991){Stecker}, {Done}, {Salamon}, \&
  {Sommers}}]{Stecker1991}
{Stecker}, F.~W., {Done}, C., {Salamon}, M.~H., \& {Sommers}, P. 1991, \prl,
  66, 2697

\bibitem[{{Stecker} {et~al.}(1992){Stecker}, {Done}, {Salamon}, \&
  {Sommers}}]{Stecker91-2}
{Stecker}, F.~W., {Done}, C., {Salamon}, M.~H., \& {Sommers}, P. 1992, \prl,
  69, 2738

\bibitem[{{Tully} {et~al.}(2008){Tully}, {Shaya}, {Karachentsev}, {Courtois},
  {Kocevski}, {Rizzi}, \& {Peel}}]{Tully08}
{Tully}, R.~B., {Shaya}, E.~J., {Karachentsev}, I.~D., {et~al.} 2008, \apj,
  676, 184

\bibitem[{{Yang} {et~al.}(2001){Yang}, {Wilson}, \& {Ferruit}}]{Yang_2001}
{Yang}, Y., {Wilson}, A.~S., \& {Ferruit}, P. 2001, \apj, 563, 124

\bibitem[{{Yuan} {et~al.}(2020){Yuan}, {Fausnaugh}, {Hoffmann}, {Macri},
  {Peterson}, {Riess}, {Bentz}, {Brown}, {Bont{\`a}}, {Davies}, {Rosa},
  {Ferrarese}, {Grier}, {Hicks}, {Onken}, {Pogge}, {Storchi-Bergmann}, \&
  {Vestergaard}}]{Yuan_distance}
{Yuan}, W., {Fausnaugh}, M.~M., {Hoffmann}, S.~L., {et~al.} 2020, \apj, 902, 26

\bibitem[{{Zanin} {et~al.}(2013){Zanin}, {Carmona}, {Sitarek}, {Colin},
  {Frantzen}, {Gaug}, {Lombardi}, {Lopez}, {Moralejo}, {Satalecka}, {Scapin},
  \& {Stamatescu}}]{Zanin13}
{Zanin}, R., {Carmona}, E., {Sitarek}, J., {et~al.} 2013, in International
  Cosmic Ray Conference, Vol.~33, International Cosmic Ray Conference, 2937

\end{thebibliography}

%
%
\begin{appendix} 

\section{Background photon field and opacity} \label{App: Low energy}

The AGN SED is constructed following the analytical prescription of \cite{Marconi_2004} \citep[see also][]{Peretti_UFO}. 
Such parametrization is characterized by a thermal component produced by the accretion disk  and an X-ray power law due to the corona. 
These two components are connected by an empirical scaling, known as $\alpha_{OX},$ which depends on the overall intrinsic AGN luminosity.
Even though it plays a subdominant role in terms of photon density for the spatial regions considered in our analysis, for the sake of completeness we also include the parametrization of the infrared peak \citep{Mullaney_2011} produced by the dusty torus surrounding the AGN. 
While, with good approximation, the X-ray and optical big blue bump originate from the innermost AGN region (corona and accretion disk), the infrared emission is spread over a much larger volume given the parsec-scale size of the dusty torus.
In agreement with \cite{Lyu21_Torus}, we assume 1 pc as a characteristic size for the torus.
\begin{figure}
    \centering
    \includegraphics[width=\columnwidth]{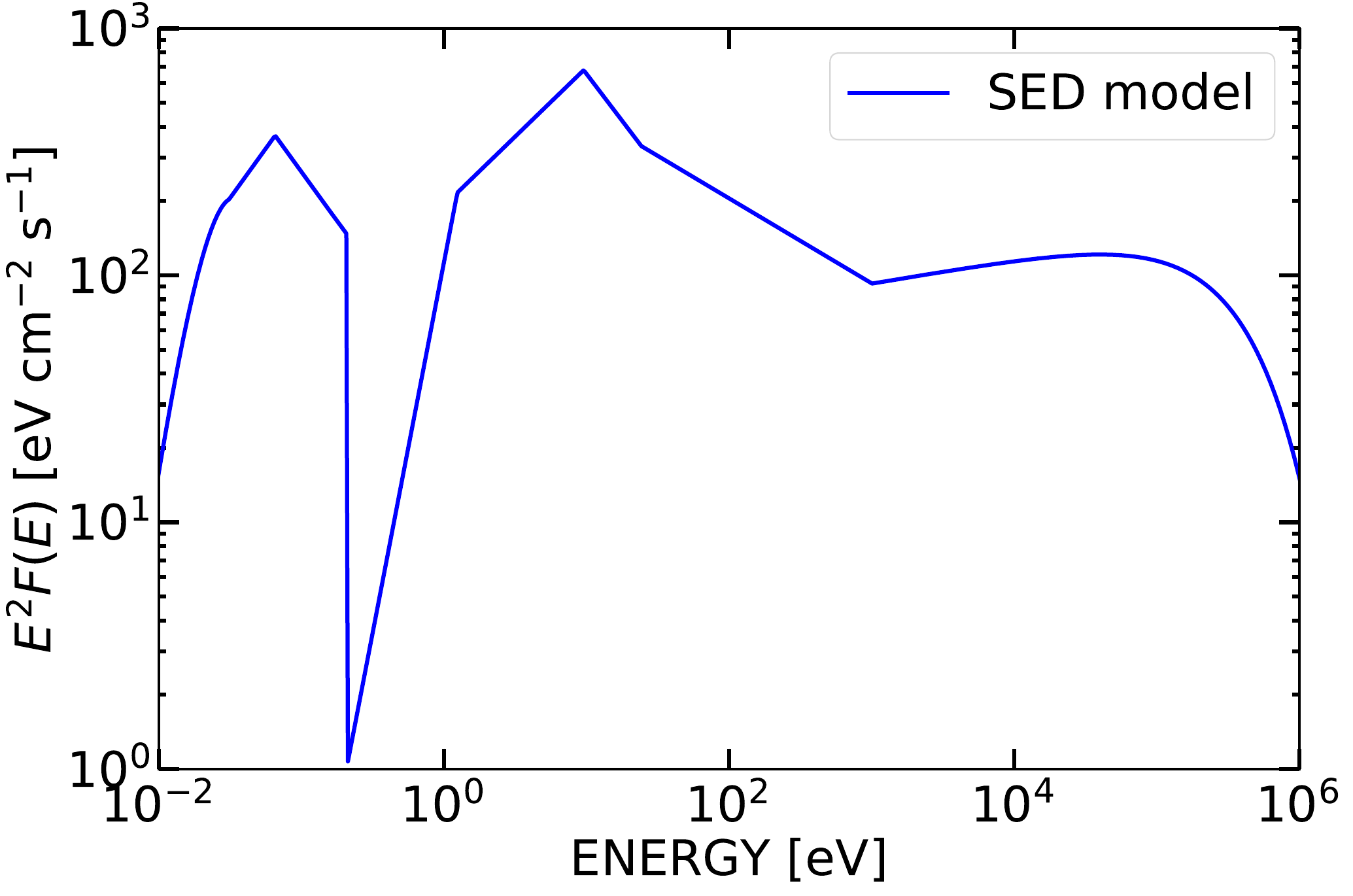}
    \caption{SED of an AGN with an X-ray luminosity $L_X = 8 \times 10^{42} \, \rm erg \, s^{-1}$.}
    \label{Fig: LE SED}
\end{figure}

Fig.~\ref{Fig: LE SED} represents the assumed  AGN intrinsic SED. 
The curve is normalized  to an X-ray luminosity of $L_X = 8 \times 10^{42} \, \rm erg \, s^{-1}$ \citep{Yang_2001,Gianolli_2023}.
This corresponds to the following bolometric luminosity $L_{\rm bol} \approx 1.9 \times 10^{44} \, \rm  erg \, s^{-1}$.

The optical depth for gamma-gamma absorption is computed following \cite{Aharonian_2008}:
\begin{equation}
\label{Eq: optical depth}
    \tau_{\gamma \gamma} (E_{\gamma}) = R \int_{\epsilon_{\rm min}}^{\infty} d\epsilon \, n_{\rm ph}(\epsilon) \sigma_{\gamma \gamma} (E_{\gamma},\epsilon)
\end{equation}
where $n_{\rm ph}$ is the background photon density evaluated at energy $\epsilon$, $\epsilon_{\rm min} = m_e^2 c^4 / E_{\gamma}$ is the energy threshold for the pair-production with a gamma-ray of energy $E_{\gamma}$.
We assume that X-ray and optical-UV photon fields are homogeneous in our region of interest of size $R$. Consequently, excluding the infrared domain, where the torus emission dominates, the photon density  scales with the size as $n_{\rm ph} \sim R^{-2}$.
For this reason, the effective optical depth scales as $\tau_{\gamma \gamma} \sim R^{-1}$.
The pair-production cross section reads \citep{Aharonian04}
\begin{align}
\label{Eq: gamma-gamma cross section}
    \sigma_{\gamma \gamma} (E_{\gamma},\epsilon) =& \frac{3 \sigma_T}{2 s^2}  \Big[\Big(s+\frac{1}{2} \ln{(s)} - \frac{1}{6}+\frac{1}{2s}\Big) \, {{\rm ln} (\sqrt{s} + \sqrt{s-1})} \nonumber \\  -& \Big(s + \frac{4}{9} - \frac{1}{9s}\Big)\Big(\sqrt{1-\frac{1}{s}}\Big)\Big],
\end{align}
where $s=\epsilon E_{\gamma}/(m_e^2c^4)$ and $\sigma_T$ is the  Thompson cross section.

\begin{figure}
    \centering
    \includegraphics[width=\columnwidth]{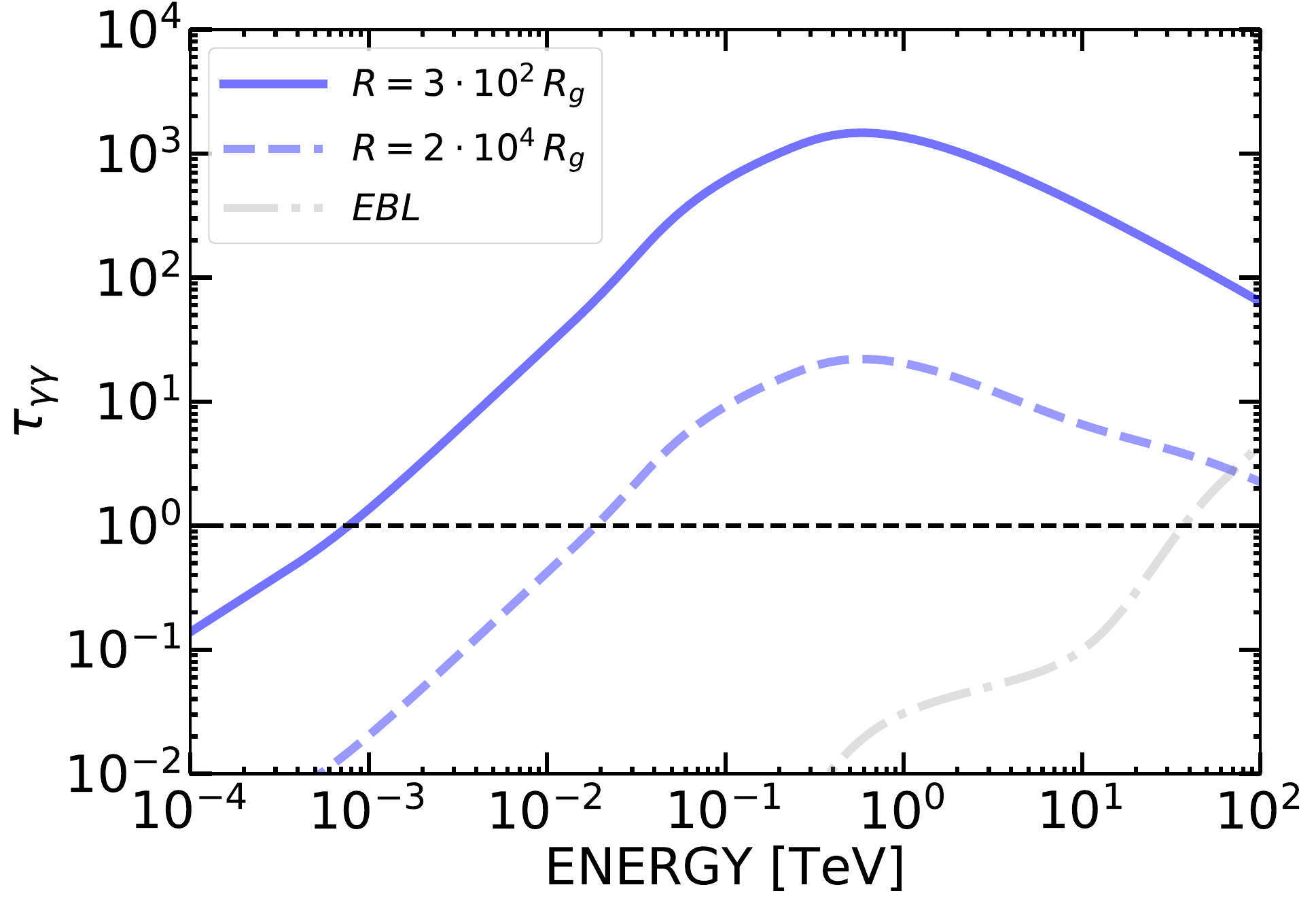}
    \caption{Optical depth for $\gamma\gamma$ absorption of gamma-ray photons from the source assuming different sizes for the various emitting regions $R$ (blue lines). The optical depth due to interaction with EBL photons \citep{Franceschini_EBL} is shown by the gray line.}
    \label{Fig: Tau_gg}
\end{figure}

Fig.~\ref{Fig: Tau_gg} illustrates the behavior of the optical depth as a function of the energy under different assumptions for the size of the emitting region. In particular, we show the optical depth corresponding to the gravitational radii assumed in calculating the absorbed gamma-ray emission described in Section \ref{Discussion}.

\end{appendix} 
\end{document}